\documentclass{aa}  

\usepackage{graphicx}
\usepackage[varg]{txfonts}
\usepackage[pdfpagelabels=false]{hyperref}	
\hypersetup{colorlinks=true,linkcolor=blue,citecolor=blue,filecolor=blue,urlcolor=blue,draft=false}
\usepackage{xspace}
\usepackage[usenames,dvipsnames,svgnames,table]{xcolor}

\newcommand{\Obelisk}{\textsc{Obelisk}\xspace}
\newcommand{\hagn}{\textsc{Horizon-AGN}\xspace}
\newcommand{\newh}{\textsc{New-Horizon}\xspace}
\newcommand{\sphinx}{\textsc{Sphinx}\xspace}
\newcommand{\ramsesrt}{\textsc{Ramses-RT}\xspace}
\newcommand{\ramses}{\textsc{Ramses}\xspace}
\newcommand{\bpass}{\textsc{Bpass}\xspace}
\newcommand{\hi}{\ion{H}{I}\xspace}
\newcommand{\hii}{\ion{H}{II}\xspace}
\newcommand{\hei}{\ion{He}{I}\xspace}
\newcommand{\heii}{\ion{He}{II}\xspace}
\newcommand{\heiii}{\ion{He}{III}\xspace}
\newcommand{\Mvir}{\ifmmode {M_{\rm vir}} \else $M_{\rm vir}$\xspace\fi}
\newcommand{\Mstar}{\ifmmode {M_\star} \else $M_{\star}$\xspace\fi}
\newcommand{\Msun}{\ifmmode {\rm M}_{\sun} \else ${\rm M}_\sun$\xspace\fi}
\newcommand{\Mbh}{\ifmmode {M_\bullet} \else $M_{\bullet}$\xspace\fi}
\newcommand{\Lbol}{\ifmmode {L_{\rm bol}} \else $L_{\rm bol}$\xspace\fi}
\newcommand{\fedd}{\ifmmode {f_{\rm Edd}} \else $f_{\rm Edd}$\xspace\fi}
\newcommand{\NH}{\ifmmode {N_{\rm H}} \else $N_{\rm H}$\xspace\fi}

\newcommand{\Rvir}{\ifmmode {R_{\rm vir}} \else $R_{\rm vir}$\xspace\fi}
\newcommand{\fesc}{\ifmmode {f_{\rm esc}} \else $f_{\rm esc}$\xspace\fi}
\newcommand{\fescstar}{\ifmmode {f_{\rm esc}^{\star}} \else $f_{\rm esc}^{\star}$\xspace\fi}
\newcommand{\fescAGN}{\ifmmode {f_{\rm esc}^{\rm AGN}} \else $f_{\rm esc}^{\rm AGN}$\xspace\fi}
\newcommand{\Mgal}{\ifmmode {{\rm M}_{1500}} \else ${\rm M}_{1500}$\xspace\fi}
\newcommand{\Magn}{\ifmmode {M_{1450}} \else $M_{1450}$\xspace\fi}
\newcommand{\MUV}{\ifmmode {{\rm M}_{\rm UV}} \else ${\rm M}_{\rm UV}$\xspace\fi}
\newcommand{\lya}{\ifmmode {{\rm Ly}\alpha} \else ${\rm Ly}\alpha$\xspace\fi}

\begin{document} 

\title{The \Obelisk simulation: Galaxies contribute more than AGN to \hi reionization of protoclusters}
\titlerunning{\Obelisk: galaxy-driven reionization of protoclusters}

\author{Maxime Trebitsch
  \inst{1,2,3}\fnmsep\thanks{\email{m.trebitsch@rug.nl}}
  \and
  Yohan Dubois\inst{1}
  \and
  Marta Volonteri\inst{1}
  \and
  Hugo Pfister\inst{1,4,5}
  \and
  Corentin Cadiou\inst{1,6}
  \and
  Harley Katz\inst{7}\fnmsep\thanks{Visitor} 
  \and
  Joakim Rosdahl\inst{8} 
  \and
  Taysun Kimm\inst{9} 
  \and
  Christophe Pichon\inst{1,10} 
  \and
  Ricarda S. Beckmann\inst{1} 
  \and
  Julien Devriendt\inst{7,8} 
  \and
  Adrianne Slyz\inst{7}
}

\institute{
  Sorbonne Universit{\'e}, CNRS, UMR 7095, Institut d'Astrophysique de Paris, 98 bis bd Arago, 75014 Paris, France
  \and
  Max-Planck-Institut f{\"u}r Astronomie, K{\"o}nigstuhl 17, 69117 Heidelberg, Germany
  \and
  Zentrum f{\"u}r Astronomie der Universität Heidelberg, Institut f{\"u}r Theoretische Astrophysik, Albert-Ueberle-Str. 2, 69120 Heidelberg, Germany
  \and
  DARK Cosmology Centre, Niels Bohr Institute, University of Copenhagen, Juliane Maries Vej 30, DK-2100 Copenhagen {\O}, Denmark
  \and
  Department of Physics, The University of Hong Kong, Pokfulam Road, Hong Kong
  \and
  Department of Physics and Astronomy, University College London, London WC1E 6BT, UK
  \and
  Sub-department of Astrophysics, University of Oxford, Keble Road, Oxford OX1 3RH, UK
  \and
  Univ Lyon, Univ Lyon 1, Ens de Lyon, CNRS, Centre de Recherche Astrophysique de Lyon, UMR5574, F-69230, Saint-Genis-Laval, France
  \and
  Department of Astronomy, Yonsei University, 50 Yonsei-ro, Seodaemun-gu, Seoul 03722, Republic of Korea
  \and
  Korea Institute for Advanced Study (KIAS), 85 Hoegiro, Dongdaemun-gu, Seoul, 02455, Republic of Korea
}

\date{Received 10 February 2020; accepted 25 May 2021}

\abstract
{
  We present the \Obelisk project, a cosmological radiation-hydrodynamics simulation that follows the assembly and reionization of a protocluster progenitor during the first two billion years after the big bang, down to $z = 3.5$.
  The simulation resolves haloes down to the atomic cooling limit and tracks the contribution of different sources of ionization: stars, active galactic nuclei, and collisions.
  The \Obelisk project is specifically designed to study the coevolution of high-redshift galaxies and quasars in an environment favouring black hole growth. In this paper, we establish the relative contribution of these two sources of radiation to reionization and their respective role in establishing and maintaining the high-redshift ionizing background.
  Our volume is typical of an overdense region of the Universe and displays star formation rate and black hole accretion rate densities similar to those of high-redshift protoclusters.
  We find that hydrogen reionization happens inside-out, is completed by $z \sim 6$ in our overdensity, and is predominantly driven by galaxies, while accreting black holes only play a role at $z \sim 4$.
}

\keywords{ Methods: numerical -- Galaxies: formation --  Galaxies: high-redshift -- intergalactic medium -- quasars: supermassive black holes -- dark ages, reionization, first stars}
   
\maketitle
%

\section{Introduction}

Observations of galaxies across cosmic time indicate that the Universe was significantly more active during its infancy compared to today. The cosmic star formation rate density (SFRD) increases rapidly with cosmic time, reaching its peak around $z \sim 2$ ('cosmic noon'). During the first three billion years from the big bang, galaxies convert their gas into stars very rapidly: The typical growth timescale of these early galaxies is two to ten times shorter than in the present Universe \citep[e.g.][]{Madau2014}. As a result of this rapid growth, approximately the same fraction of the stellar mass observed today was formed before $z > 2$ and after $z \sim 0.7$.
As they quickly form stars, the first galaxies illuminate the initially neutral intergalactic medium (IGM). Accreting massive black holes (BHs) harboured in some of these galaxies also shine as they grow in mass.
These sources of radiation start ionizing their environment from their formation at $z \gtrsim 20$, so that by $z \sim 6$ ('cosmic dawn'), almost all of the hydrogen in the Universe is (re)ionized \citep[e.g.][]{Fan2006b}; a similar process happens for helium at a later time, with \heii reionization being complete around $z \sim 2.5-3.5$ \citep[e.g.][]{Shull2010,Worseck2016}. 

There has been tremendous progress in recent years in observing those sources of reionization: A large number of galaxies has now been found at $z \geq 6$ \citep[see e.g. the review of][]{Stark2016}, with some candidates even reaching $z \sim 10-11$, when the Universe was less than 500 Myr old \citep{Oesch2016, Salmon2018, Lam2019, Bouwens2019}. Yet, how the assembly of these early galaxies results in the large-scale reionization of the Universe in less than one billion years is still largely an open problem. For instance, while we know that high-redshift galaxies produce ionizing photons very efficiently, the fraction, \fesc, of these ionizing photons that manage to escape the interstellar (ISM) and circumgalactic medium (CGM), and hence contribute to reionization, is still largely unconstrained. Significant observational effort has been undertaken to measure \fesc both at high \citep[e.g.][]{Mostardi2015, Shapley2016, Grazian2017, Naidu2018, Vanzella2016, Vanzella2018, Steidel2018, Fletcher2019, Tanvir2019} and low-redshift \citep{Leitet2011, Leitet2013, Borthakur2014, Leitherer2016, Izotov2016a, Izotov2016b, Izotov2018a, Izotov2018b}, and yet we still seem to be only scratching the surface. The lack of constraints on \fesc limits our understanding of the epoch of reionization (EoR) as changing the value of \fesc by a factor of two drastically affects the timing of reionization \citep[e.g.][]{Madau2017, Dayal2018}.

Moreover, the relative role of the different ionizing sources -- star forming galaxies and active galactic nuclei: AGN -- in reionizing the Universe is one of the most pressing questions of the field \citep[e.g.][]{Madau1999, Haehnelt2001, Robertson2015, Haardt2015, Madau2015, Garaldi2019}, and the need for sources beyond star forming galaxies depends strongly on the assumed value for \fesc \citep[e.g.][etc.]{Yoshiura2017, Finkelstein2019, Naidu2020, Dayal2020}. While bright quasars such as those found at $z > 7$ by \citet{Mortlock2011} or \citet{Banados2018} are too rare to be dominant actors of reionization \citep[e.g.][but see \citet{Giallongo2019} for a different perspective]{Becker2013}, they have been suggested \citep[e.g.][]{Chardin2015} as a solution to explain the patchiness of the late stages of reionization \citep[e.g.][]{Becker2015}.
Revealing the detailed properties of these distant galaxies and BHs to build a consistent picture of the EoR is a key science project for the next generation of observatories such as the \emph{James Webb Space Telescope} or the \emph{Square Kilometre Array}.

In the standard $\Lambda$ cold dark matter ($\Lambda$CDM) cosmological model, structures form hierarchically, meaning that low-mass structures form first: This implies that at a given time the most massive structures will on average be older. As a consequence, galaxies found in the most massive haloes have a mass assembly history even more skewed towards early times. For instance, models such as those of \citet{Behroozi2013} predict that for galaxies found in today's clusters, living in dark matter (DM) haloes with virial masses of order $\Mvir \gtrsim 10^{14}\,\Msun$, most of the stellar mass was assembled before $z \gtrsim 2$.

Because of their accelerated evolution \citep{Overzier2016}, protoclusters are unique laboratories for studying the complex processes governing galaxy formation and evolution in the high-redshift Universe, a crucial step towards understanding the assembly history of galaxies in our local Universe.
Additionally, following the growth of a protocluster can prove extremely useful for studying the accretion of gas onto galaxies. In the \citet{Birnboim2003} and \citet{Dekel2006} picture, accretion onto these haloes should transition from a `cold mode' to a `cold in hot' mode and finally reach the regime where cosmological filaments penetrating the halo are shock-heated and destroyed in the hot atmosphere; this transition is expected to set the scene for the subsequent quenching of galaxies.

The most massive of these protoclusters are expected to harbour galaxies that host some of the most massive BHs, with masses in excess of $\Mbh \gtrsim 10^9\,\Msun$ already at $z \sim 6$. The accretion onto these BHs results in feedback phenomena releasing copious amounts of energy into the surrounding environment \citep[e.g.][]{Fabian2012}, which has been suggested as a viable mechanism to explain the coevolution of supermassive BHs and their host galaxies \citep[e.g.][]{Silk1998}: Understanding to which extent this AGN feedback operates is one of the  main challenges in the field of galaxy formation today. Here again, protoclusters could be very useful probes of this phenomenon: Compared to the field, these objects are found to be richer in AGN than average environments \citep[e.g.][]{Casey2016}.

On the theoretical side, tremendous effort has been made in the recent years to improve cosmological simulation models. Large simulation projects, such as \hagn \citep{Dubois2014}, \textsc{Eagle} \citep{Crain2015, Schaye2015}, \textsc{Illustris} \citep{Vogelsberger2014} and its sucessor \textsc{Illustris-TNG} \citep{Marinacci2018, Naiman2018, Nelson2018, Pillepich2018b, Springel2018}, or \textsc{MassiveBlack-II} \citep{Khandai2015} and its high-redshift successor \textsc{BlueTides} \citep{Feng2016}, have been designed to study the evolution of the galaxy population in a cosmological volume. These simulations have been very successful at reproducing and explaining, for example, the diversity of galaxies in the low-redshift Universe. Motivated by this success, several teams have been developing a new generation of simulations that reach a much higher resolution while keeping the large-scale cosmological environment, for instance \textsc{Illustris-TNG50} \citep{Nelson2019,Pillepich2019}, \textsc{Romulus} \citep{Tremmel2017}, or \newh \citep{Dubois2020,Park2019}.

Because of the additional complexity introduced by the inclusion of radiation hydrodynamics and the resulting extra computational cost, the landscape of cosmological simulations attempting to model reionization self-consistently is more sparsely populated. On top of this, radiative transfer strongly couples the hierarchy of scales involved in galaxy formation: while reionization is a global process that needs to be modelled on scales larger than $\gtrsim 200\,h^{-1}$ comoving Mpc (cMpc)  to sample cosmic variance \citep{Iliev2014}, the physical mechanisms affecting the source properties operate at the scale of the ISM \citep[e.g.][]{Kimm2014, Paardekooper2015}, with some studies even suggesting that simulations need to resolve molecular clouds to include all relevant processes \citep{Howard2018,Kimm2019,Kim2019}. At the same time, the inhomogeneous reionization can act as a feedback loop and affect galaxy formation, for instance by suppressing star formation in low-mass haloes \citep[e.g.][]{Shapiro1994, Shapiro2004, Gnedin2000, Katz2020} or around rare bright sources.

Several projects, such as the \textsc{Croc} project \citep{Gnedin2014}, the \textsc{CoDa} project \citep{Ocvirk2016,Ocvirk2020}, the \textsc{Aurora} simulation \citep{Pawlik2017}, and the \textsc{Technicolor-Dawn} simulation \citep{Finlator2018}, have taken the approach of modelling a large volume and including the sources with subgrid models. These have provided a very fruitful approach for studying the evolution of the IGM and of the broad galaxy population at $z \gtrsim 6$. The other approach, taken for instance by the \textsc{Renaissance} suite simulations \citep{OShea2015} and the \sphinx project \citep{Rosdahl2018}, has been to carry out full volume simulations with a resolution comparable to zoom simulations ($\lesssim 10\,\mbox{pc}$), at the cost of strongly reducing the simulated volume. This has led to major advances in the modelling of the ISM of the galaxies responsible for reionizing the Universe. Because of the limited volume, these simulations can only model average\footnote{An exception to this is the \textsc{Renaissance}-\emph{Rare peak} simulation, but it has only been run until $z \sim 12$.} environments that are dominated by low mass galaxies. Importantly, these simulations all focused on the study of galaxies during the EoR and are therefore terminated by $z \sim 6$.

\begin{figure*}
  \centering
  \includegraphics[width=0.85\linewidth]{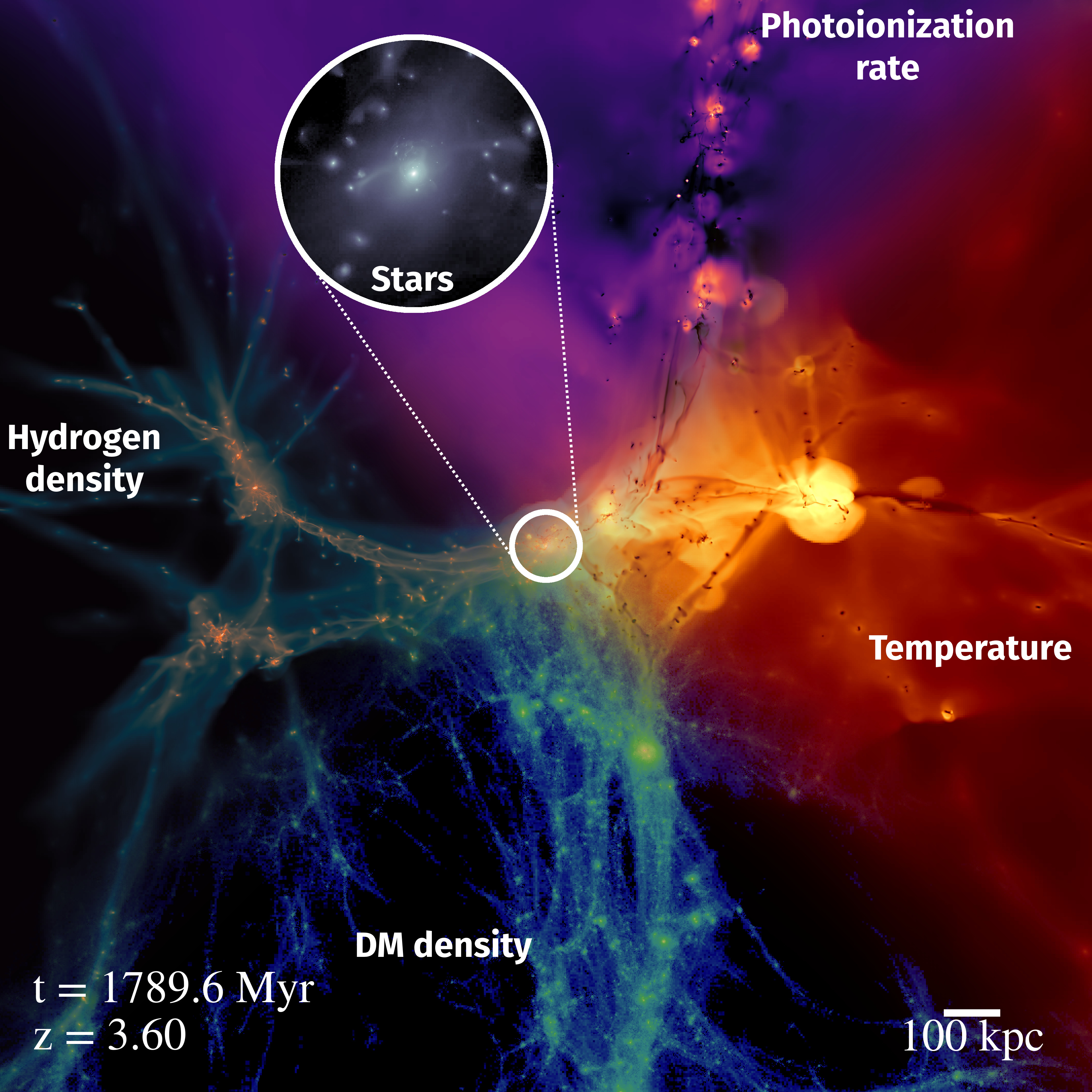}
  \caption{Snapshot of the central region of the \Obelisk, illustrating the physics modelled in the simulation. The complex gas distribution is shown on the left, the corresponding DM skeleton on the bottom, the gas temperature on the right highlighting self-shielded filaments (dark brown) and hot feedback bubbles (in yellow), and the upper part shows the \hi photoionization rate with the knots of the cosmic web lit up by bright sources. The inset zooms in on the stellar distribution around the central galaxy.}
  \label{fig:Obelisk_highlight}
\end{figure*}
In an ideal world, one would want to bridge the gap between `galaxy evolution'-oriented simulations (\textsc{Eagle}, \hagn, \textsc{Illustris}, etc.) and the `reionization'-oriented simulations, in order to build a full picture of galaxy formation and evolution at high-redshift, during and after reionization. This is, however, numerically too challenging even for the largest simulations to date. Nevertheless, it is desirable to complement existing reionization simulations that do not model AGN and, conversely, to study the effect of radiation feedback on galaxy formation at high resolution in this mass regime. At $z = 4$, the UV luminosity function of \citet{Ono2018} suggests that AGN start to dominate at magnitudes around -23 and number densities of several $10^{-6}\,\mbox{mag}^{-1}\mbox{Mpc}^3$. We would therefore expect a few of these objects, which are not the rarest quasars in the Universe, in the \hagn volume at $z=4$. In this work, we present our attempt at capturing the physics of these galaxies and AGN through the \Obelisk project: a simulation designed to study the high-redshift Universe, following self-consistently the evolution of the largest protocluster of the \hagn volume down to $z \sim 3.5$.
This is an intermediate approach in many respects: While we do not simulate a large cosmological volume, we still capture the formation of a rare structure, unattainable in small boxes, and we retain a higher resolution than simulations such as \hagn while doing so. At the same time, the connection to the larger \hagn simulation allows us to track the descendants of the \Obelisk galaxies to connect the high and low-redshift galaxy populations.
The project focuses on the full high-redshift evolution, going beyond the end of the EoR at $z \sim 6$, and is unique in that it follows the build up and maintenance of the ionizing background using both galaxies and AGN as sources.

This paper, the first of a series based on the exploitation of the \Obelisk simulation, is structured as follows: we begin in Sect.~\ref{sec:methods} with a comprehensive description of our numerical methodology. In Sect.~\ref{sec:populations}, we present the galaxy and BH populations emerging from the \Obelisk model, and in Sect.~\ref{sec:reionization} discuss the relative role of these populations in reionizing the volume.
We finally present a summary of our results in Sect.~\ref{sec:summary}


\section{The \Obelisk simulation}
\label{sec:methods}

\begin{figure*}
  \centering
  \resizebox{\hsize}{!}{
    \includegraphics{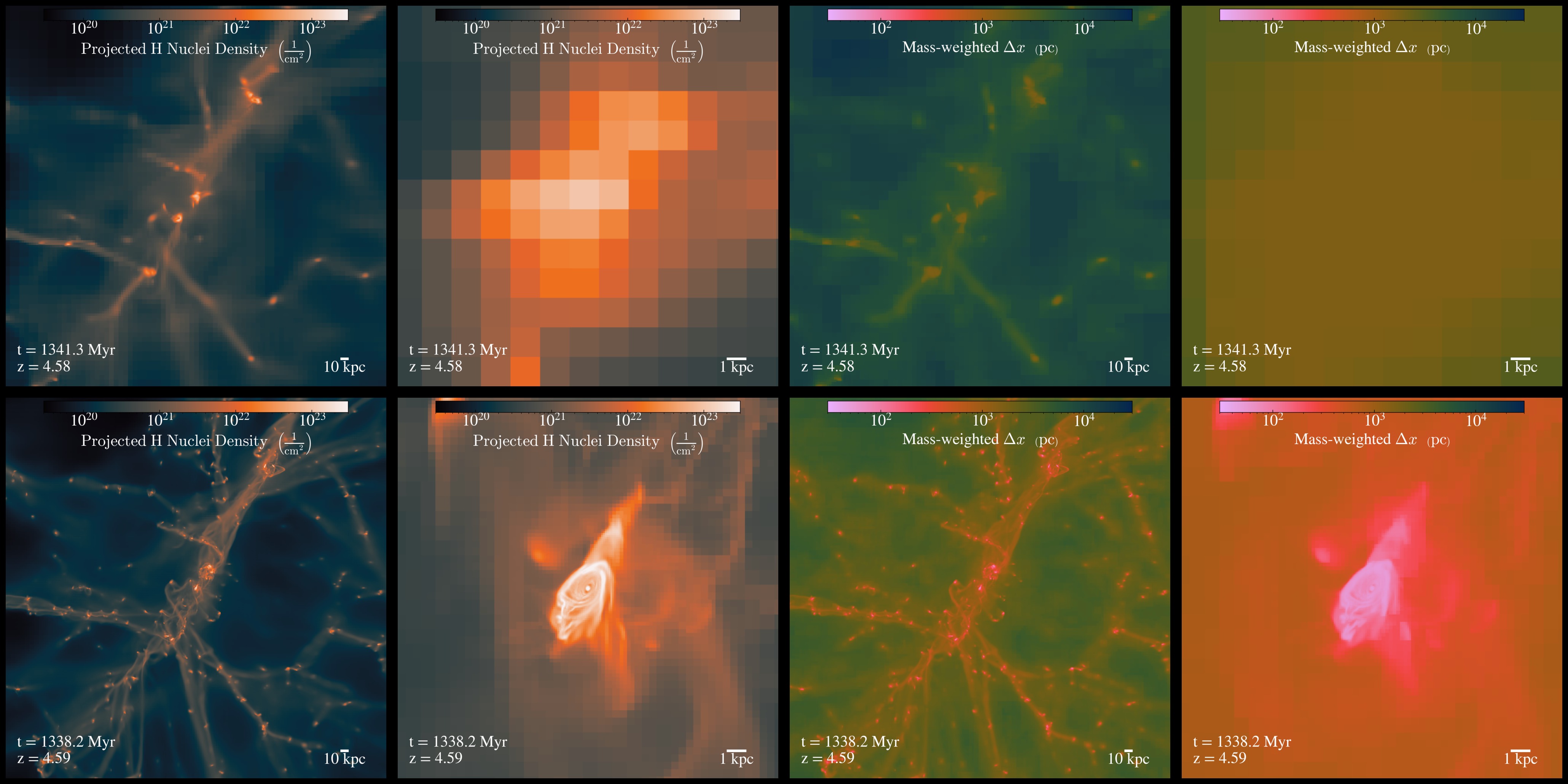}
  }
  \caption{Illustration of the increased resolution between \hagn (upper row) and \Obelisk (lower row) for the projected gas density (first two columns) and typical cell size (last two columns). The first and third columns highlight the large-scale environment, where the resolution in the filaments goes from $\Delta x_{\hagn} \sim 4\,\mbox{kpc}$ to $\Delta x_{\Obelisk} \lesssim 500\,\mbox{pc}$, and the second and fourth columns shows the improvement at the galaxy scale.}
  \label{fig:HAGN_vs_Obelisk}
\end{figure*}

In this paper, we introduce the \Obelisk simulation, a high-resolution ($\Delta x \simeq 35\,\mbox{pc}$), radiation-hydrodynamical simulation of a sub-volume of \hagn\footnote{\label{fn:HAGN}\url{https://www.horizon-simulation.org/}} \citep{Dubois2014}. We describe in this section the code and physical models that we use in the simulation: the initial conditions and volume selection (Sect.~\ref{sec:ICs}), the broad features of our radiation hydrodynamics simulation code (Sect.~\ref{sec:ramsesRT}), and the physical models we employ for stars (Sect.~\ref{sec:subgrid-star}), BHs (Sect.~\ref{sec:subgrid-bh}), and dust physics (Sect.~\ref{sec:dust}). We describe our halo and galaxy identification strategy in Sect.~\ref{sec:halo-galaxy-ident}.

\subsection{Initial conditions and volume selection}
\label{sec:ICs}

The initial conditions for the \Obelisk simulation were chosen to follow the high-redshift evolution of an overdense environment. For this purpose, we chose to re-simulate with a high resolution the region around the most massive halo at $z\sim 2$ in the \hagn simulation: Selecting initial conditions based on this simulation allows our results to be compared and contrasted with the work that has already resulted from \hagn.

The \hagn simulation follows the evolution of a cosmological volume of side $L_{\rm box} = 100\, h^{-1}\,\mbox{cMpc}$ (and periodic boundary conditions), assuming a $\Lambda$CDM cosmology compatible with the 7-year data from the \emph{Wilkinson Microwave Anistropy Probe} \citep{Komatsu2011}: Hubble constant $H_0 = 70.4\,\mbox{km}\,\mbox{s}^{-1}\,\mbox{Mpc}^{-1}$, total matter density $\Omega_{\rm m} = 0.272$, dark energy density $\Omega_{\Lambda} = 0.728$, baryon density $\Omega_{\rm b} = 0.0455$, amplitude of the matter power spectrum $\sigma_8 = 0.81$, and scalar spectral index $n_s = 0.967$.
The initial conditions have been created with \textsc{MPgrafic} \citep{Prunet2008,Prunet2013} with $1024^3$ DM particles and as many gas cells (corresponding to a DM mass resolution of $M_{\rm DM,LR} = 8\times 10^7\,\Msun$). In the original \hagn run, the grid is then adaptively refined over the course of the simulation, maintaining a maximum spatial resolution of $\Delta x = 1$ proper kpc at all redshifts down to $z=0$. We improve on this resolution by a factor of $\sim 30$ in our \Obelisk sub-volume, reaching a resolution of $\Delta x = 35\,\mbox{pc}$ (see Sect.~\ref{sec:hydrograv}).

In order to achieve our target resolution at a reasonable computation cost, we only re-simulated a fraction of the initial \hagn volume. To this end, we selected the most massive halo (of virial mass is around $\Mvir \simeq 2.5 \times 10^{13}\,\Msun$) in the simulation at $z \sim 2$ as identified by the \textsc{AdaptaHOP} halo finder \citep{Aubert2004,Tweed2009}. By $z = 0$, this halo remains the most massive cluster of \hagn, with $\Mvir \sim 6.6 \times 10^{14}\,\Msun$.
We first identified all the particles within $4\Rvir$ of the target halo at $z = 1.97$, namely in a sphere of radius $2.51\,h^{-1} \mbox{cMpc}$ and tracked them back to the initial conditions. We computed the convex hull enclosing all these particles in the initial conditions and defined this region as our high-resolution patch.
We then created a re-sampled version of the \hagn initial conditions with $4096^3$ DM particles, corresponding to a mass resolution of $M_{\rm DM,HR} = 1.2\times 10^6\,\Msun$.
Finally, we selected all the high-resolution particles belonging to the patch previously defined, and embedded this patch in the larger \hagn box, using successively lower and lower resolution regions as buffers, until reaching an effective resolution of $256^3$ particles in the outer parts of the volume.
We filled the full volume with a passive variable whose value is $1$ within the high-resolution patch and $0$ outside, and used this as a refinement mask (see Sect.~\ref{sec:hydrograv} for further details).
Fig.~\ref{fig:HAGN_vs_Obelisk} illustrates the gain in resolution between \hagn (upper row) and \Obelisk (lower row).

Finally, the \Obelisk simulation improves upon its parent \hagn in several important ways (beyond resolution) which we describe in detail in Sect.~\ref{sec:ramsesRT}, \ref{sec:subgrid-star} and \ref{sec:subgrid-bh}.
We should note at this point that a similar methodology has been employed for the \newh simulation, which focuses on an average region of the Universe. Apart from the radiation-hydrodynamical evolution, our numerical methodology is kept as close as possible to the \newh simulation, so as to facilitate the comparison between the two simulations.

\subsection{Halo and galaxy identification}
\label{sec:halo-galaxy-ident}
We identified galaxies and haloes in each snapshot of the simulation using the \textsc{AdaptaHOP} halo finder \citep{Aubert2004, Tweed2009} with the most massive sub-maximum method (MSM) to separate between host haloes and substructures.
In this framework, haloes and subhaloes are groups of particles located at maxima of the density field, and the MSM method requires that the most massive sub-structure is defined as the central object.
Compared to previous works using \textsc{AdaptaHOP} \citep[e.g.][for \hagn]{Dubois2014}, we amended the halo finder to identify structures using all collisionless particles, both stars and DM. The DM halo is then identified to the DM component of the (sub-)structure, while the galaxy is defined as all stellar particles in the (sub-)structure. In the following, we refer interchangeably to (sub-)structures and (sub-)haloes when discussing groups produced by our modified \textsc{AdaptaHOP}.
To be elected as a candidate halo, a structure has to exceed a threshold density $\rho_{\rm t}$. Instead of using a fixed density (e.g. 200 times the average or critical density), we used the fit from \citet{Bryan1998}, yielding an density of roughly $\rho_{\rm t} \lesssim 178$ for the redshift interval studied here. We only considered structures with more than $n_{\rm members} \geq 100$ particles. Notably, we only considered in the analysis galaxies with more than 100 star particles. This yields $52\,428$ ($69\,235$) host haloes,  $41\,244$ ($116\,663$) subhaloes and $12\,549$ ($67\,478$) galaxies at $z = 6.0$ ($z = 3.53$), respectively.

Once a (sub-)halo has been identified, we fit a tri-axial ellipsoid to it, and we find the largest ellipsoid for which the virial theorem is verified. We used this ellipsoid to define the virial radius \Rvir, and the virial mass \Mvir is the mass enclosed in this ellipsoid. In addition to properties global to the total (DM + stellar) structure, we also measured several quantities separately for the stars and DM component, such as the half-mass radius $R_{50}$. For the galaxy, we also computed additional kinematical and morphological informations such as the projected effective radius $R_{\rm eff}$, the star formation rate, or the mass-weighted age and metallicity.
We note that a galaxy can correspond to a sub-structure and still lie outside of the virial radius of its parent structure. While we did not separate the populations of central and satellite galaxies in this work, it is worth emphasizing that not all galaxies in sub-structures will correspond to satellite galaxies.

\subsection{Radiation hydrodynamics with \ramsesrt}
\label{sec:ramsesRT}

The \Obelisk simulation was run with \ramsesrt \citep{Rosdahl2013,Rosdahl2015}, a multi-group radiative transfer (RT) extension of the public, adaptive mesh refinement (AMR) code \ramses\footnote{\label{fn:ramses}\url{https://bitbucket.org/rteyssie/ramses/}} \citep{Teyssier2002}. \ramses follows the evolution of DM, gas, stars, and BHs, via gravity, hydrodynamics, radiative transfer, and non-equilibrium thermochemistry.

\subsubsection{Hydrodynamics and gravity}
\label{sec:hydrograv}

The gas was evolved using an unsplit second-order MUSCL-Hancock scheme \citep{vanLeer1979}, based on the Harten-Lax-van Leer-Contact (HLLC) Riemann solver \citep{Toro1994} to solve the Euler equations. A MinMod total variation diminishing scheme was used to reconstruct the inter-cell conservative variables from the cell-centred values. We assumed an ideal gas equation of state with adiabatic index $\gamma = 5/3$ to close the relation between internal energy and gas pressure. 
Gravity was modelled by projecting the collisionless particles (stars and DM) onto the AMR grid using cloud-in-cell interpolation and solving the Poisson equation using the multigrid particle-mesh method described in \citet{Guillet2011} on coarse levels and conjugate gradient on fine levels with a transition at level $\ell = 13$.

In the high-resolution region, the initial mesh was refined up to a spatial resolution of $\Delta x \simeq 35\,\mbox{ckpc}$ (equivalent to $4096^3$, or an initial grid level $\ell_{\rm min,HR} = 12$), and a passive refinement scalar was set to a value of $1$ within that region. We only allowed refinement where the value of this refinement scalar exceeds $0.01$, effectively ensuring that only the initial high-resolution region was adaptively refined throughout its collapse.
Within this region, we allowed for ten extra levels of refinement, up to a maximal spatial resolution of $\Delta x \simeq 35\,\mbox{pc}$ varying within a factor of two depending on the redshift.
Our refinement criterion follows the standard \ramses quasi-Lagrangian approach: A cell is selected for refinement if $\rho_{\rm DM} \Delta x^3 + (\Omega_{\rm DM}/\Omega_b)\rho_{\rm gas} \Delta x^3+ (\Omega_{\rm DM}/\Omega_b) \rho_* \Delta x^3 > 8\ M_{\rm DM,HR}$, where $\rho_{\rm DM}$, $\rho_{\rm gas}$ and $\rho_*$ are the DM, gas and stellar densities in the cell, respectively. In a DM-only run, this would refine a cell as soon as it contains at least eight high-resolution DM particles.
We note that cells hosting sink particles and the associated clouds (see Sect.~\ref{sec:subgrid-bh}) are always maximally refined.
In order to keep the physical resolution constant over the course of the simulation (the box has a constant comoving size), we only permitted a new level of refinement when the expansion scale factor doubles (in our case, $a_{\rm exp} = 0.1$ and $0.2$). While this is known to induce a small temporary increase in the star formation \citep[e.g.][]{Snaith2018}, it ensures that the physical subgrid models that have been derived with a specific physical resolution in mind are always used at the appropriate scale.

While the baryonic mass (from gas, stars, and BHs) is directly projected onto the maximally refined grid, we smoothed the DM density field by depositing the mass of the DM particles on a coarser grid ($\Delta x \simeq 540\,\mbox{pc}$).
This level of smoothing corresponds to the maximal level of refinement triggered in an analogue DM-only simulation where no absolute maximal level of refinement was enforced. This ensures that the effective size of the DM particles correspond to their mass resolution.

The Courant-Friedrichs-Lewy stability condition was enforced using a Courant factor of 0.8, even though the duration of the timestep is predominantly set by the radiation solver (see Sect.~\ref{sec:radiation}).

\subsubsection{Radiation}
\label{sec:radiation}

The details of the methods used for the injection, propagation, and interaction of the radiation with hydrogen and helium are described in \citet{Rosdahl2013}, and we therefore only summarize the main features here.

The RT module propagates the radiation emitted by massive stars and accreting BHs (see Sect.~\ref{sec:stellar-radiation} and \ref{sec:bh-radiation} for details on the source models) in three frequency intervals describing the \hi, \hei, and \heii photon fields (between $13.6 - 24.59\,\mbox{eV}$, $24.59 - 54.4\,\mbox{eV}$, and $54.4 - 1000\,\mbox{eV}$, respectively).
The first two moments of the equation of RT are solved on the AMR grid using an explicit first-order Godunov method with the M1 closure \citep{Levermore1984, Dubroca1999} for the Eddington tensor. The radiation is then coupled to the hydrodynamical evolution of the gas through the non-equilibrium thermochemistry for hydrogen and helium and radiation pressure (Sect.~\ref{sec:non-equil-therm}).

As we used an explicit solver, we are subject to a Courant-like condition for the propagation of the radiation. This is an extremely stringent condition for the radiation: As the speed of light is much larger than any other velocity in the simulation, the RT timestep should in principle be extremely short. We mitigated this in two ways, following a similar approach to the \sphinx simulation \citep{Rosdahl2018}.
First, subcycled the RT timestep on each AMR level, with up to 500 RT steps for each hydro step, while preventing photons from crossing level boundaries during the subcycling (see the discussion in Sect.~2.4 of \citealt{Rosdahl2018}). In addition to this, we used the traditional approach of artificially reducing the speed of light by a constant factor, $f_c$, to prevent too short a timestep and too large a number of RT subcycles. This `reduced speed of light' approximation, initially proposed by \citet{Gnedin2001} and used here following the implementation of \citet{Rosdahl2013}, works well when studying the ISM and the CGM of individual galaxies, where the propagation of light is effectively limited by the propagation of ionization fronts.
Contrary to single galaxy or small volume simulations performed with \ramsesrt \citep[e.g.][]{Kimm2017, Trebitsch2017, Costa2019}, the \Obelisk simulation tracks the reionization process in a reasonably large volume of the Universe (typically $\gtrsim 10^4\,\mbox{Mpc}^3$ comoving at $z \sim 4$): Because of this we {can} no longer fully employ this `reduced speed of light' approximation.
We instead used the so-called `variable speed of light' approximation \citep[VSLA, ][]{Katz2017}, where the factor $f_c$ is local. Ideally, one would have a relatively low speed of light in the densest regions and a higher speed of light in voids. With our quasi-Lagrangian refinement strategy, this can be achieved by tying the reduction factor to the cell size; with a cell twice as small as its neighbour will use a value of $f_c$ reduced by a factor of two. Following \citet{Katz2018} and the \sphinx simulation, we used $f_c = 0.2$ for the coarsest cells in the high-resolution region ($\ell = 12$ or $\Delta x \simeq 35\,\mbox{ckpc}$) as the reionization history should be fairly converged with this value \citep[e.g.][]{Deparis2019,Ocvirk2019}. We then chose to decrease $f_c$ by a factor of two per level until $f_c = 0.0125$ at all levels above $\ell \geq 16$ ($\Delta x \simeq 2\,\mbox{kpc}$). For the low resolution cells outside of the high-resolution region, we set the speed of light to a low value ($f_c = 0.01$). While this creates some accumulation of radiation just outside of the region of interest, the effects on the high-resolution region are negligible. 

As one of the major endeavours of the \Obelisk project is to study the relative contributions of massive stars and of AGN to the establishment and maintenance of the ionizing UV background, we used the photon tracer algorithm of \citet{Katz2018,Katz2019c} to keep track of the contribution of different sources to the local photon flux (and hence the strength of the UV background) and ionization of the gas. In a nutshell, we track the fractional contribution of each type of source to the radiation density and flux when the photons are injected and advected on the grid. With this, we also keep track of the fraction of the gas that has been photoionized by each type of source, and the difference with the total \hii fraction in each cell is the fraction of the gas that has been collisionally ionized.
This allows us, for example, to track across cosmic time which sources contribute the most to \hi-photoionization rate in the IGM, which sources are responsible for ionizing which environment, and to compute population-averaged escape fractions. This algorithm has already been applied in \citet{Katz2018} and \citet{Katz2019} to study the contribution to reionization of stellar sources of different ages or mass, or residing in haloes of difference masses.
In this work, we traced photons based on their original source: We explicitly follow the radiation produced by stellar populations and by accreting BHs.
We refer the interested reader to the Sect.~2 of \citet{Katz2018} for details on the implementation of the algorithm.

\subsubsection{Gas thermochemistry}
\label{sec:non-equil-therm}
\ramsesrt features non-equilibrium thermochemistry by following the ionization state of hydrogen and helium: \element{H}, \element[+]{H}, \element{He}, \element[+]{He}, \element[++]{He}, as described in \citet{Rosdahl2013}. 
The coupling with the radiation is performed via photoionization, collisional ionization, collisional excitation, recombination, bremsstrahlung, homogeneous cooling and heating off the cosmic microwave background, and di-electronic recombinations. 
The whole thermochemistry step is subcycled within every RT step, and uses the smooth injection approach from \citet{Rosdahl2013} to limit the amount of subcycling. We further assume the on-the-spot approximation: Any ionizing photon emitted by recombination is assumed to be absorbed locally, and we thus ignore emission of ionizing radiation resulting from direct recombinations to the ground level.

We include a cooling contribution from metals using the standard approach in \ramses. Above $T \geq 10^4\,\mbox{K}$, we use the cooling rates computed with \textsc{Cloudy} \footnote{\label{fn:cloudy}\url{http://www.nublado.org/}} \citep[][version 6.02]{Ferland1998} assuming photoionization equilibrium with the redshift-dependent \citet{Haardt1996} UV background. We stress that we do not use this UV background for the hydrogen and helium non-equilibrium photo-chemistry, but solely for computing the metal cooling rates.
We also account for energy losses via metal line cooling below $T \leq 10^4\,\mbox{K}$, following the prescription of \citet{Rosen1995} based on \cite{Dalgarno1972}, and approximate the effect of the metallicity by scaling the metal cooling enhancement linearly (we still assume a Solar-like abundance pattern for simplicity).
This allows the gas to cool down to a temperature floor of $T_{\rm floor} = 50\,\mbox{K}$. We chose to use a density-independent temperature floor rather than a (density-dependent) pressure floor, usually used to prevent numerical fragmentation \citep{Truelove1998}, because our model for star formation (see the next section) is constructed to efficiently remove gas to form stars in regions with high density and low temperature (which would be susceptible to numerical fragmentation).

Because we do not include molecular hydrogen, we adopted a homogeneous initial metallicity floor of $Z = 10^{-3} Z_\odot$ in the whole computational volume, to allow the gas to cool down below $10^4\,\mbox{K}$ before the formation of the first stars (and the subsequent metal enrichment). Aside from this, the gas in the box is initially neutral and composed of $X=76\,\%$ hydrogen and $X=24\,\%$ helium by mass.

\subsection{Stellar populations}
\label{sec:subgrid-star}

We model stars as particles with mass $m_\star \simeq 10^4\,\Msun$ representing a single stellar population, assuming a fully sampled \citet{Kroupa2001} initial mass function (IMF) between $0.1$ and $100\,\Msun$.

\subsubsection{Star formation}
\label{sec:star-formation}
We only consider cells to be star forming when the local number density $n_{\rm gas}$ (or equivalently the mass density $\rho_{\rm gas}$) exceeds a threshold $n_{\rm SF} = 5\,\mbox{H}\,\mbox{cm}^{-3}$ (chosen as the typical ISM density), and when the local turbulent Mach number defined as the ratio of the turbulent velocity to the sound speed exceeds $\mathcal{M} \geq 2$.
The amount of gas converted into stars follows a \citet{Schmidt1959} law:
\begin{equation}
  \label{eq:sfr_schmidt}
  \dot{\rho}_{\star} = \epsilon_{\star} \frac{\rho_{\rm gas}}{t_{\rm ff}},
\end{equation}
so that on average $M_{\rm sf} = \dot{\rho}_\star \Delta x^3 \Delta t = \epsilon_{\star} \rho_{\rm gas} \Delta x^3 \Delta t/t_{\rm ff}$ of gas is converted into star particles during one timestep $\Delta t$, where $G$ is the gravitational constant, $\epsilon_{\star}$ is the local star formation efficiency per free-fall time and $t_{\rm ff} = \sqrt{3\pi / 32 G \rho_{\rm gas}}$ is the gas free-fall time.

One key difference between the method we use here, as already discussed for example in \citealt{Trebitsch2017} or \citealt{Kimm2017}, and the traditional approach used for instance in \hagn is that the star formation efficiency $\epsilon_{\star}$ is a local parameter, rather than a constant, and depends on the local gas density $\rho_{\rm gas}$, sound speed $c_{\rm s}$, and turbulent velocity $\sigma_{\rm gas}$. Here, we approximated $\sigma_{\rm gas}$ by taking the velocity differences between the host cell and its immediate neighbours.
The analytic expression for $\epsilon_{\star}(\rho_{\rm gas}, c_s, \sigma_{\rm gas})$ follows the `multi-ff PN' model of \citet{Federrath2012, Padoan2011}:
\begin{equation}
  \label{eq:sfr_ff}
  \epsilon_{\star} \propto \text{e}^{\frac{3}{8}\sigma_s^2}\left[1 + \mathrm{erf}\left(\frac{\sigma_s^2 - s_{\rm crit}}{\sqrt{2\sigma_s^2}}\right)\right],
\end{equation}
for $\mathcal{M} \geq 2$ and $0$ otherwise, and where $\sigma_s = \sigma_s(\sigma_{\rm gas}, c_{\rm s})$ characterizes the turbulent density fluctuations, $s_{\rm crit} = \mathrm{ln}(\rho_{\rm gas, crit}/\rho_{\rm gas})$ is the critical density above which the gas will be accreted onto stars, and $\rho_{\rm gas, crit} \propto (\sigma_{\rm gas}^2 + c_{\rm s}^2) \sigma_{\rm gas}^2/c_{\rm s}^2$. In practice, this means that $\epsilon_\star$ increases with $\sigma_{\rm gas}$, and when the virial parameter decreases.

The actual number $N$ of particles formed in one cell in one timestep $\Delta t$ is drawn from a Poisson distribution $P(N) = (\lambda^N/N!) \exp(-\lambda)$ of parameter $\lambda = M_{\rm sf} / m_\star$ \citep[see][for details]{Rasera2006}. Additionally, for numerical stability, we limit the number of star particles formed in one timestep so that the amount of gas removed from a cell is capped at $90\%$ of the gas mass in that cell.

\subsubsection{Supernova feedback}
\label{sec:supernova-feedback}
When a star particle reaches an age of $t_{\rm SN} = 5\,\mbox{Myr}$, we assume that a mass fraction $\eta_{\rm SN} = 0.2$ of the initial stellar population explodes as SNe and returns mass and metals to its environment with a yield of $0.075$. Each SN explosion releases an energy $E_{\rm SN} = 10^{51}\,\mbox{erg}$ assuming an average progenitor mass of $m_{\rm SN} \simeq 20\,\Msun$, corresponding to an average of 1 SN explosion for $100\,\Msun$ of stars. At our mass resolution, each star particle releases $10^{53}\,\mbox{erg}$ per event, instantaneously.

The explosion itself was modelled following the mechanical feedback implementation of \citet{Kimm2014, Kimm2015}, which injects radial momentum according to the phase of the explosion (energy conserving or momentum conserving) that is resolved. We refer the reader to these works for the details of the algorithm implementation, and once again only recall the main features here.
If we resolve the adiabatic expansion phase of the SN, we directly inject the kinetic energy and momentum corresponding to $10^{51}\,\mbox{erg}$ to the cells neighbouring the SN site. If this energy conserving phase is not resolved, the SN explosion is only captured in its final snowplough phase; in this case we directly inject the terminal momentum $p_{\rm snow}$ in the neighbouring cells. We determine the phase of the SN explosion that we resolved on a cell-by-cell basis: For each of the cells neighbouring the SN site, we evaluate the mass ratio $\chi_{\rm SN} = \mathrm{d}M_{\rm swept} / \mathrm{d}M_{\rm ej}$ between the swept material (ejecta + swept ISM) and the ejecta, and compare it to a critical mass ratio $\chi_{\rm SN,tr}$.
For low values of $\chi_{\rm SN}$ (i.e. in the adiabatic phase), we inject
\begin{equation}
  \label{eq:psn_ad}
  \Delta p_{\rm ad} = f_{\rm geom} \sqrt{2\chi_{\rm SN} M_{\rm ej} f_{\rm e} E_{\rm SN}}\,,
\end{equation}
where $f_{\rm geom}$ is a geometrical factor describing how the total energy and mass of the SN is split between the neighbouring cells, and $f_{\rm e} = 1 - (\chi_{\rm SN} -1)/(3\chi_{\rm SN,tr} -1)$ ensures a smooth transition between the two modes of momentum injection.
In the snowplough phase, we inject the terminal momentum \citep[e.g.][]{Thornton1998, Blondin1998, Kim2015, Martizzi2015}
\begin{equation}
  \label{eq:psn_snow}
  \Delta p_{\rm snow} = 3\times 10^{5}\,\Msun\,\mbox{km}\,\mbox{s}^{-1}\, f_{\rm geom} E_{\rm SN,51}^{16/17} n_{\element{H}}^{-2/17} \tilde{Z}^{-0.14}\,,
\end{equation}
where $E_{\rm SN,51}$ is the total SN energy in units of $10^{51} \,\mbox{erg}$, $n_{\element{H}}$ is the local hydrogen number density in units of $\mbox{cm}^{-3}$, and $\tilde{Z}$ is the local metallicity in units of $Z_\odot$ floored at $0.01\, Z_\odot$.

We further assumed that, the photo-ionization pre-processing of the ISM by young OB stars prior to the SN event augments the final radial momentum from a SN \citep{Geen2015}. While this should be taken into account by the radiative transfer in our simulation, \citet{Trebitsch2017} argued that a significant fraction of the ionizing radiation can be emitted in regions where the Str{\"o}mgren radius, $r_S$, is not resolved in which case this momentum increase will often be missed. Thus, as in \citet{Trebitsch2018} and \citet{Rosdahl2018}, we followed the subgrid model of \citet{Kimm2017}, which adds this missing momentum when the Str{\"o}mgren radius is locally not resolved, when $r_S < \Delta x$. We then increased the pre-factor in Eq.~\ref{eq:psn_snow} to $5 \times 10^{5}\,\Msun\,\mbox{km}\,\mbox{s}^{-1}$ following the results of \citet{Geen2015}.

\subsubsection{Stellar radiation}
\label{sec:stellar-radiation}
Using the \sphinx simulation, \citet[][]{Rosdahl2018} have shown that the inclusion of binary stars has a strong effect on the timing of reionization because binary interactions lead to an increased and more sustained production of ionizing radiation \citep[e.g.][see also \citealt{Topping2015}]{Stanway2016, Gotberg2019}.
We emit ionizing radiation from each star particle following the spectral energy distribution (SED) resulting from the Binary Population and Spectral Synthesis code v2.2.1 \citep[\bpass,][\emph{Tuatara} version]{Eldridge2017, Stanway2018}, which includes the effect of these binary interactions. In the \emph{Tuatara} release, the binary fraction of stars depends on the initial stellar mass, with around 60\% (6\%) of low-mass (high-mass) stars being isolated.
More specifically, we used the model \texttt{imf135\_100} closest to a \citet{Kroupa2001} IMF with slopes $-1.30$ between $0.1 - 0.5\,\Msun$ and $-2.35$ between $0.5 - 100\,\Msun$.
Each star particle is assigned a luminosity in each radiation bin as a function of its age and metallicity, scaled directly with the mass of the particle. As described in \citet{Rosdahl2013}, the average energy of each radiation bin and the interaction cross-sections are updated every five coarse timesteps, so that they accurately represent the properties of the average source population.

While lower resolution simulations usually include a subgrid correction to account for the unresolved escape fraction and calibrated to reproduce reasonable reionization history \citep[e.g.][]{Gnedin2014, Ocvirk2016, Ocvirk2020, Pawlik2017, Finlator2018}, we followed here the approach taken by simulations reaching a spatial resolution better than a few tens of parsecs and use the luminosity of the star particle directly. We stress that we do not claim that $\Delta x \simeq 35\,\mbox{pc}$ is sufficient to resolve in great detail the rich multi-phase structure of the ISM; we acknowledge that the uncertainty on the ISM structure could affect the escape of ionizing radiation from birth clouds either way: Unresolved ionized channels could lead to a higher escape fraction \citep[e.g.][]{Ma2015}, while unresolved clumping could increase the amount of absorption in the ISM \citep[e.g.][]{Yoo2020}. We note however that the tests performed in the \sphinx framework \citep[appendix B of][]{Rosdahl2018} suggest that a resolution of $\Delta x \simeq 20\,\mbox{pc}$ yields galaxy properties similar to their fiducial run with $\Delta x \simeq 10\,\mbox{pc}$
Finally, we note that the absence of subgrid calibration of the escape fraction is consistent with the assumptions we made for the star formation model and for the BH accretion (see Sect.~\ref{sec:subgrid-bh}): namely, that we broadly resolve the large-scale ISM density distribution and the formation of molecular clouds. Changing one ingredient (e.g. the unresolved escape of radiation) without changing the others would break that consistency.

\subsection{BH model}
\label{sec:subgrid-bh}

We followed the same approach to model BHs, based on the implementation of \citet{Dubois2010, Dubois2012,Dubois2014c}: BHs are modelled as sink particles, which are allowed to accrete gas and release energy, momentum and radiation into their environment. We shall now describe the various aspects of our BH model.

\subsubsection{BH formation}
\label{sec:bh-formation}
We seeded the sink particles representing our BHs with an initial mass $M_{\bullet,0} = 3 \times 10^4\,\Msun$ in cells where the following criteria are met: the gas density $n_{\rm gas}$ and stellar density $n_\star$ must both locally exceed a threshold $n_{\rm sink} = 100\,\mbox{H}\,\mbox{cm}^{-3}$; and the gas must be Jeans-unstable.
We imposed an exclusion radius $r_{\rm excl} = 50\,\mbox{kpc}$ to avoid the formation of multiple BHs in the same galaxy.
Each sink particle is dressed with a swarm of `cloud' particles, positioned on a regular grid lattice within a sphere of radius $4 \Delta x$ and equally spaced by $\Delta x/2$.
These cloud particles provide a convenient way of probing and averaging the properties of the gas around the BH.
At our resolution, this means we probe a sphere of radius $4\Delta x \simeq 140\,\mbox{pc}$ around each BH with 2109 clouds.
We set the initial velocity of the BH to that of its host cell, and assigned it a spin $a = 0$.

Our seed mass choice stems from physical as well as numerical considerations. The BH formation mechanism is highly uncertain, with different models predicting very different seed masses, from $\sim 10\,\Msun$ to $\gtrsim 10^5\,\Msun$ \citep[e.g.][]{Volonteri2010}. Numerically, having a BH seed less massive than the mass of star particles is not desirable: \citet{Pfister2019} suggests that otherwise, the BH trajectory becomes extremely sensitive to the fluctuation of the (stellar) gravitational potential. As our mass resolution for star particles is $m_\star \simeq 10^4\,\Msun$, we choose a BH seed mass three times higher.
The underlying physical picture would be that of a light seed BH that has already undergone some early growth or of a heavier seed, and is consistent with our choice of seeding BHs only in regions with a sufficiently high stellar density (thus mimicking pre-galactic centres).

\subsubsection{BH accretion}
\label{sec:accretion}
Once a sink particle has formed, it grows through two channels: BH-BH mergers and gas accretion.
We allowed two BHs to merge when they are closer than $4\Delta x$ from one another, and only if their relative velocity is lower than the escape velocity of the binary system they would form in vacuum.
As we are far from resolving the gaseous accretion disc around BHs, we used the classical Bondi-Hoyle-Lyttleton \citep{Bondi1952} approach to compute the accretion rate onto the BH:
\begin{equation}
\dot{M}_{\rm BHL} = 4\pi G^2 \Mbh^2 \frac{\bar{\rho}}{\left(\bar{c}_s^2 + \bar{v}_{\rm rel}^2\right)^{3/2}}\,,
\label{eq:accrate-bhl}
\end{equation}
where $\Mbh$ is the BH mass, $\bar{\rho}$ and $\bar{c}_s$ are respectively the average gas density and sound speed, and $\bar{v}_{\rm rel}$ is the relative velocity between the BH and the surrounding gas. The bar notation denotes an averaging over the cloud particles using a Gaussian kernel $w \propto \exp\left(-r^2/r_{\rm sink}^2\right)$, where $r_{\rm sink}$ is defined using the Bondi radius $r_{\rm BHL} = \frac{G\Mbh}{c_s^2 + v_{\rm rel}^2}$:
\begin{equation}
  \label{eq:rsink}
  r_{\rm sink} = 
  \begin{cases}
    \Delta x/4 & \text{if}\quad r_{\rm BHL} < \Delta x/4, \\
    r_{\rm BHL}  & \text{if}\quad \Delta x/4 \leq r_{\rm BHL} < 2 \Delta x, \\
    2 \Delta x & \text{if}\quad 2 \Delta x \leq r_{\rm BHL}.
  \end{cases}
\end{equation}
We did not use any extra artificial boost for the gas accretion onto the BH. The accretion rate was capped at the Eddington rate:
\begin{equation}
\dot{M}_{\rm Edd} = \frac{4\pi G \Mbh m_p}{\epsilon_r \sigma_{\rm T} c},
\label{eq:accrate-edd}
\end{equation}
where $m_p$ is the proton mass, $\sigma_{\rm T}$ is the Thompson cross section, $c$ is the speed of light and $\epsilon_r$ is the radiative efficiency of the accretion flow onto the BH, which depends on the spin of the BH (see Sect.~\ref{sec:bh-spin}). Additionally, only a fraction $1-\epsilon_r$ of the mass accreted onto the accretion disc effectively reaches the BH: The rest is radiated away. At very low accretion rates, when $\chi = \dot{M}_{\rm BHL}/\dot{M}_{\rm Edd}$ drops below $\chi_{\rm crit} = 0.01$, the flow becomes radiatively inefficient, in which case we followed \citet[][Eq.~16]{Benson2009} and reduce the radiative efficiency of the flow to $\tilde{\epsilon_{r}} = \epsilon_r (\chi/\chi_{\rm crit})$.
The final BH accretion rate is therefore:
\begin{equation}
  \label{eq:accrate-both}
  \dot{\Mbh} = (1-\tilde{\epsilon_r}) \min\left(\dot{M}_{\rm BHL}, \dot{M}_{\rm Edd}\right).
\end{equation}
Gas is then removed from cells within $4 \Delta x$ of the sink particle in a kernel-weighted fashion, and the accretion is capped to prevent the BH from removing more than 25\% of the gas content of the cell in one timestep for numerical stability (i.e. $w \dot{\Mbh}\Delta t \lesssim 0.25 \rho_{\rm gas}\Delta x^3$).

\subsubsection{BH dynamics}
\label{sec:bh-dynamics}
While the dynamics of the sink particles is computed at the most refined grid level (see Sect.~\ref{sec:hydrograv}), we lack the resolution to accurately capture the effects of dynamical friction that will affect the detailed BH dynamics. For instance, \citet{Pfister2017} showed with a resolution study that resolving the influence radius of a BH is crucial to get a chance to resolve the formation of BH binaries in the aftermath of a galaxy merger; this length-scale being of the order of $1\,\mbox{pc}$ for a BH with mass $\Mbh \sim 10^6\,\Msun$ in a Milky-Way-like galaxy, and much lower for less massive BHs, it is well below the resolution that modern cosmological simulations such as \Obelisk can reach.
While most of these simulations resort to moving the sink particle towards a local minimum of the potential \citep[e.g.][]{Crain2015,Pillepich2018a,Dave2019}, we took here a different approach and used a subgrid model to account for the effect of the unresolved dynamical friction \citep[see also][]{Tremmel2015}.

We used the drag force implementation introduced in \citet{Dubois2013} to model the force exerted by the gaseous wake lagging behind the BH.
The frictional force has an analytic expression given by \citet{Ostriker1999}, and is proportional to $F_{\rm DF} = \alpha f_{\rm gas} 4\pi \rho (G \Mbh / \bar{c_s}^2)$, where $\alpha$ is an artificial boost, with $\alpha  = (\rho/\rho_{\rm DF, th})^2$ if $\rho > \rho_{\rm th}$ and 1 otherwise, and $f_{\rm gas}$ is a fudge factor varying between 0 and 2 which depends on the BH Mach number $\mathcal{M}_\bullet = \bar{v}_{\rm rel}/\bar{c}_s$ \citep{Ostriker1999, Chapon2013}.
In light of the results of \citet{Beckmann2018} who showed that this subgrid model begins to fail when the wake is resolved, we set $F_{\rm DF} = 0$ whenever the influence radius $2\mathrm{G} M_\bullet / \max(\bar{c_s},\bar{v_\mathrm{rel}})^2 > 0.2\Delta x$.
In this work, we took $\rho_{\rm DF, th} \simeq 0.003\ \mbox{to}\ 0.01 \,\mbox{cm}^{-3}$. We discuss the effect of this ad hoc choice in Appendix~\ref{sec:dfthreshold}.

We also included a contribution to the dynamical friction caused by collisionless particles (stars and DM separately), analogous to what happens to the gas: A gravitational wake of stars and DM is created by the passage of a massive body (the BH) and will decelerate it \citep{Chandrasekhar1943, Binney1987}.
Our implementation, described in detail in \citet{Pfister2019}, is somewhat similar to that of \citet{Tremmel2015} used in the \textsc{Romulus} simulation \citep{Tremmel2017}.
We directly compute the contribution of the collisionless particles within a sphere of radius $4\Delta x$.
The deceleration is parallel to the velocity $\mathbf{v}_\bullet$ of the BH relative to the background (stars and DM) and has a magnitude $a_{\rm DF}$ is computed as follow:
\begin{align}
  \label{eq:aDF-chandra}
  a_{\rm DF} = -4\pi \frac{G^2 \Mbh}{v_\bullet^2} &\left( \log\Lambda  \int_0^{v_\bullet} 4\pi v_\bullet^2 f(v) \mathrm{d}v \right.\nonumber\\
   & + \left. \int_{v_\bullet}^\infty 4\pi v^2 f(v) \left[\log\left(\frac{v+v_\bullet}{v-v_\bullet}\right) - 2\frac{v_\bullet}{v}\right] \mathrm{d}v \right)\,,
\end{align}
where $v_\bullet$ is the norm of the relative velocity with respect to the mass-weighted velocity of the surrounding collisionless particles, $\Lambda = 4\Delta x / r_{\rm def}$ is the Coulomb logarithm with $r_{\rm def} = G\Mbh/v_\bullet^2$, and $f$ is the distribution function defined with the velocities of the particles withing the sphere of radius $4\Delta x$. As for the gas, we switch off the subgrid model when the influence radius is resolved by more than $0.2\Delta x$
\begin{equation}
  \label{eq:distribfunction}
  4\pi v^2 f(v) = \frac{3}{256\pi \Delta x^3}\sum_i m_i \delta(v_i - v).
\end{equation}
We refer the interested reader to \citet{Pfister2019} for a more detailed discussion of the model.

\subsubsection{BH spin evolution}
\label{sec:bh-spin}
We follow self-consistently the evolution of the spin magnitude and direction over the course of the simulation using the implementation presented in \citet[][see also \citealt{Fiacconi2018,Bustamante2019} for similar implementations]{Dubois2014b, Dubois2014c}. Here again, we refer the interested reader to these works for an extensive discussion of the model and tests of its validity.

We evolve the magnitude of the spin following gas accretion through an expression derived in \citet{Bardeen1970}:
\begin{equation}
  \label{eq:spinevol-bardeen}
  a^{n+1} = \frac{1}{3} \frac{r_{\rm isco}^{1/2}}{M_{\rm ratio}} \left[4 - \left(3\frac{r_{\rm isco}}{M_{\rm ratio}^2} - 2\right)^{1/2}\right],
\end{equation}
where $M_{\rm ratio} = \Mbh_{,n+1}/\Mbh_{,n}$ ($\Mbh_{,n}$ being the mass of the BH at times $t_n$), and
\begin{equation}
  \label{eq:risco}
  r_{\rm isco} = R_{\rm isco}/R_{\rm g} = 3 + Z_2 \mp \operatorname{sign}(a) \sqrt{(3-Z_1)(3+Z_1+2Z_2)}
\end{equation}
is the radius of the innermost stable circular orbit (ISCO) expressed in units of gravitational radius $R_{\rm g}$. $Z_1$ and $Z_2$ are function of the spin magnitude $a$, given by
\begin{align}
  \label{eq:spinZ1}
  \begin{aligned}
    Z_1 &= 1 + (1-a^2)^{1/3}\left[(1+a)^{1/3} + (1-a)^{1/3}\right] \\
    Z_2 &= \left(3a^2 + Z_1^2\right)^{1/2}.
\end{aligned}
\end{align}
The $\mp$ sign in Eq.~\ref{eq:risco} depends on whether the BH is co-rotating ($a \geq 0$) or counter-rotating ($a \leq 0$) with its accretion disc. For a co-rotating BH, $1 \leq r_{\rm isco} \leq 6$, while $6 \leq r_{\rm isco} \leq 9$ for the counter-rotation case ($r_{\rm isco} = 6$ only for a non-spinning BH).
The direction of the BH spin is evolved assuming that the angular momentum of the (unresolved) accretion disc aligns with that of the accreted gas. The potential mis-alignment between the BH spin and the accretion disc leads to the formation of a warped disc in the innermost regions of the accretion disc precessing about the spin axis because of the Lense-Thirring effect. This warped disc will eventually completely align or anti-align with the BH spin. The anti-alignment configuration occurs when the angle $\theta$ between angular momenta of the disc $\mathbf{J_{\rm d}}$ and of the BH $\mathbf{J_{\bullet}}$ fulfils $\cos\theta < -0.5 \|\mathbf{J_{\rm d}}\| / \|\mathbf{J_{\bullet}}\|$ \citep{King2005}. We assume that the accretion disc is well described by the \citet{Shakura1973} thin disc, and define $J_{\rm d}$ as the value of the angular at the smallest radius between the warp radius and the self-gravity radius.
We refer the reader to Sect.~3 of \citet{Dubois2014c} for the equations governing the details of this process, but we wish to stress here that we do not enforce the spin of the BH to always be aligned with the angular momentum of the accreted gas.

Our model assumes a thin disc solution: We only apply it at high accretion rates, when $\chi \gtrsim 0.01$.
At lower accretion rate, when the accretion flow is radiatively inefficient, we modify our model following the results of the simulations of `magnetically choked' accretion flows from \citet{McKinney2012}. In practice, we assume that each low accretion rate event fills an accretion disc, and over the course of the accretion event, the BH spin is evolved at a rate given by a `spin-up' parameter (or rather spin-down parameter, as in this regime, the absolute value of the spin magnitude tends to decrease systematically) given by a fourth-order polynomial fit of the results in Table~7 of \citet{McKinney2012}, particularly their simulations \texttt{AaaaN100} where \texttt{aaa} is the BH spin of each model. 
In any case, we limit the spin-up process to $|a| \leq a_{\rm max} = 0.998$ following \citet{Thorne1974}.
Finally, when two BHs merge, we update the spin of the remnant using the fit of \citet{Rezzolla2008}, according the measured pre-merger BH spins, orbital angular momentum, and mass ratio.

The spin evolution is not a purely passive quantity in \Obelisk: The radiative efficiency $\epsilon_r$ of the accretion flow is effectively set by the spin through $r_{\rm isco}$:
\begin{equation}
  \label{eq:effrad}
  \epsilon_r = 1 - \sqrt{1 - \frac{2}{3r_{\rm isco}}}.
\end{equation}
For a non-rotating BH, this leads to $\epsilon_r \simeq 0.057$, and the canonical $\epsilon_r = 0.1$ corresponds to $a \simeq 0.7$. 

\subsubsection{AGN feedback}
\label{sec:agn-feedback}

We implemented AGN feedback following accretion events using the dual mode approach of \citet{Dubois2012}.
At low Eddington ratio $\chi < \chi_{\rm crit}$, the AGN is in `radio mode', while it is in `quasar mode' when $\lambda_{\rm Edd} \geq 0.01$.

For the quasar mode, each sink particle dumps an amount $\dot{E}_{\rm AGN} \Delta t$ of thermal energy over a timestep $\Delta t$ in a sphere of radius $\Delta x$ centred on the BH.
The energy injection rate is calculated as
\begin{equation}
  \label{eq:Edot_quasar}
  \dot{E}_{\rm AGN} = \epsilon_f \epsilon_r \dot{M}_\bullet c^2,
\end{equation}
where $\epsilon_f = 0.15$ is the fraction of the bolometric luminosity that is transferred to the gas \citep{Booth2009,Dubois2012} , driving unresolved winds through Compton heating, and UV and IR momentum coupling of radiation. This ad hoc choice efficiency allows for the self-regulation of BH growth through their feedback.
The actual input of (UV) photons from the AGN is also treated explicitly on resolved scales with the \ramsesrt solver (see Sect.~\ref{sec:bh-radiation}).

For the radio mode, we deposit momentum as a bipolar cylindrical outflow mimicking the propagation of a jet in the surrounding gas with a velocity $u_{\rm J} = 10^4\,\mbox{km}\,\mbox{s}^{-1}$. The outflow  profile corresponds to a cylinder of radius $\Delta x$ and height $2\Delta x$ weighted by a kernel function
\begin{equation}
  \label{eq:jetprofile}
  \psi\left(r_{\rm cyl}\right) = \frac{1}{2\pi\Delta x^2}\exp\left(-\frac{r_{\rm cyl}^2}{\Delta x^2}\right),
\end{equation}
with $r_{\rm cyl}$ the cylindrical radius. The outflow removes mass from the central cell and transport it to the cells enclosed by the jet at a rate $\dot{M}_{\rm J}$ with
\begin{equation}
  \label{eq:jetmass}
  \dot{M}_{\rm J}(r_{\rm cyl}) = \frac{\psi(r_{\rm cyl})}{\Psi} \eta_{\rm J} \dot{M}_\bullet\,,
\end{equation}
where $\Psi$ is the integral of $\psi$ over the cylinder and $\eta_{\rm J} = 100$ is the mass-loading factor of the jet accounting for the interaction between the jet and the ISM at unresolved scales. The momentum is injected in a direction parallel to the BH spin (with opposite signs above and below the sink) with the norm of the momentum $q(r_{\rm cyl})$ given by
\begin{equation}
  \label{eq:jetmomentum}
  q_{\rm J}(r_{\rm cyl}) = \dot{M}_{\rm J}(r_{\rm cyl}) u_{\rm J}.
\end{equation}
We inject the corresponding kinetic energy into the gas:
\begin{equation}
  \label{eq:jetegy}
  \dot{E}_{\rm J}(r_{\rm cyl}) = \frac{q_{\rm J}(r_{\rm cyl})^2}{2 \dot{M}_{\rm J}(r_{\rm cyl})} = \frac{\psi(r_{\rm cyl})}{\Psi} \dot{E}_{\rm AGN}.
\end{equation}
Here, the injection rate $\dot{E}_{\rm AGN}$ is given by
\begin{equation}
  \label{eq:Edot_radio}
  \dot{E}_{\rm AGN} = \epsilon_{\rm MCAF} \dot{M}_\bullet c^2,
\end{equation}
where $\epsilon_{\rm MCAF}$ is given by a fourth-order polynomial fit to the simulations of \citet{McKinney2012}. More precisely, we use the same set of simulations as for the spin evolution (see Sect.~\ref{sec:bh-spin}), and sum the contributions of winds and jet based on Table~5: $\epsilon_{\rm MCAF} = \epsilon_{\rm j} + \epsilon_{\rm w,o}$ (using respectively $\eta_{\rm j}$ and $\eta_{\rm w,o}$ in their notations).

\subsubsection{BH radiation}
\label{sec:bh-radiation}
In addition to the thermal and kinetic energy injection described in the previous section, we release ionizing energy radiation from the BHs to represent the contribution of AGN to the ionizing radiation field. For this, we applied the method presented in \citet{Trebitsch2019}.
We release radiation at each fine timestep, and the amount of radiation released in each frequency bin is given by the luminosity of the quasar in each band. In this section, we highlight the main aspects of our AGN SED model, and leave the details to Appendix~\ref{sec:agn-sed-model}.

While the implementation of the photon injection is similar to the work of  \citet{Bieri2017}, it differs in the spectrum assumed for the AGN. Indeed, instead of a constant SED inspired by the averaged spectrum of \citet{Sazonov2004}, we modelled the SED of the radiation produced by the accretion onto the BH as a multi-colour black-body spectrum corresponding to a \citet{Shakura1973} thin disc, and extend it at high energy with a power-law $\alpha_{\rm UV} = -1.5$, consistent with the value derived by \citet{Lusso2015} for a sample of high-redshift quasars. We then approximated the whole spectrum with a piecewise power-law for simplicity.
We assume that $f_{\rm IR} = 30\%$ (consistent with the \citealt{Sazonov2004} spectrum) of the bolometric luminosity of the disc is absorbed by dust and re-emitted as IR radiation (which will participate to the quasar mode feedback), which we do not model here, thus leaving a total luminosity $L_{\rm rad} = 0.7 \epsilon_r \dot{\Mbh} c^2$ available for the radiation. 
To get the AGN luminosity in each frequency bin, we integrate the resulting SED in each frequency interval. Similarly, we compute the average photon energy in each bin, as well as the energy-weighted and photon-weighted interaction cross sections (see \citealt{Rosdahl2013} for details on the role of these quantities). We do not include the presence of X-rays, which can create a diffuse shell of partially ionized gas in the outskirts of cosmological \hii regions \citep{Graziani2018}.

As the disc profile is a function of $\Mbh$, $\dot{\Mbh}$ and $a$, the multi-colour black-body spectrum of the AGN will also depend on these parameters. To limit the computational cost, we pre-calculate all the quantities that depend on the AGN SED, and only interpolate between these values over the course of the simulation. Because the shape of the spectrum is weakly sensitive to the value of the spin parameter, we adopt the shape corresponding to $a = 0$. We have ensured that the adopted AGN SED yields an average spectrum similar to the AGN SED used in \citet{Volonteri2017} for a population of growing BHs at $z \sim 6$.

Finally, we note again that the thin disc solution assumes that the accretion flow is radiatively efficient: We therefore only release radiation when $\chi \geq \chi_{\rm crit}$. 
\begin{figure}
  \centering
  \includegraphics[width=\linewidth]{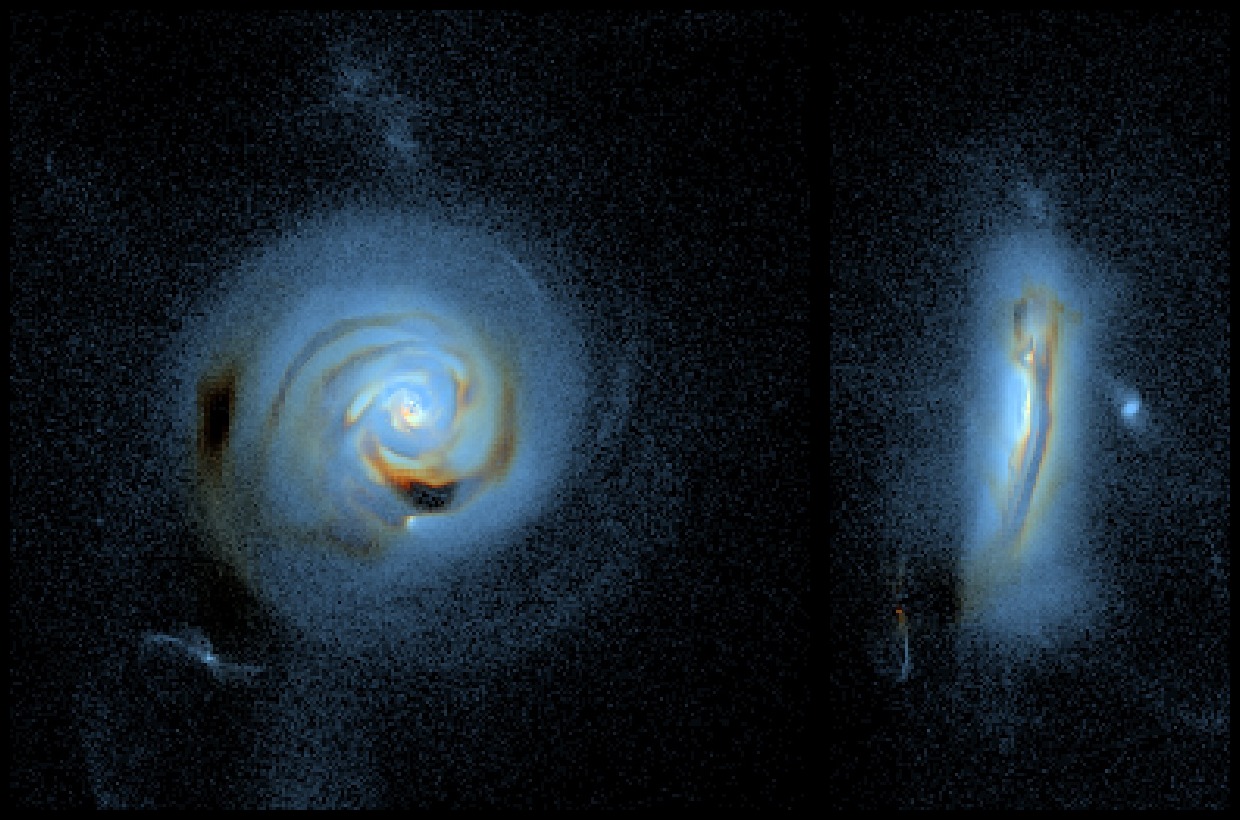}
  \caption{Mock (rest-frame) \emph{ugr} image of a $\Mstar = 4\times 10^{10}\,\Msun$ galaxy at $z = 6$, accounting for the dust attenuation along the line of sight. The almost edge-on panel on the right shows a clear dust lane.}
  \label{fig:galdust}
\end{figure}

\subsection{Dust model}
\label{sec:dust}
We included in \Obelisk a subgrid model for the evolution of dust treated as a separate constituent to that of metals locked in the gas phase. The details of the model will be described in a future work {Dubois et al. (in prep)}, and we present here the main features. Our model assumes that dust grains are released in the ISM via SN explosions, grow in mass via accretion of gas-phase metals, and are destroyed by SN explosions and via thermal sputtering. Fig.~\ref{fig:galdust} shows the resulting dust attenuation on a mock (rest-frame) \emph{ugr} image of a $z\sim 6$ galaxy with mass $\Mstar = 4\times 10^{10}\,\Msun$.

Specifically, we consider that all dust grains belong to one single population and are perfectly coupled to the gas (no dust drift relative to gas), so that they can be described with only one scalar $D$ describing the local dust mass fraction (this is similar to the approach taken by e.g. \citealt{McKinnon2017} and \citealt{Li2019}). This scalar is passively advected just as the metallicity $Z$, now representing the total metal mass fraction (in the gas phase as well as locked in the dust).
We further neglect the size distribution of grains and assume that all grains have a unique size $a_{\rm d}$. While the size distribution could in principle be followed in the model \citep[e.g.][]{McKinnon2018}, this would come at a substantial memory overhead.
In practice, we assumed that the grains have an average size of $a_{\rm g} = 0.1\,\mu\mbox{m}$ and density $\mu_{\rm g} = 2.4\,\mbox{g}\,\mbox{cm}^{-3}$.
Finally, we stress that the dust is purely passive in our simulation: While in reality, a fraction $D/Z$ of the metals is locked into dust, we do not account for this when estimating the contribution of metal cooling.

When supernovae release metals in the ISM, we assumed that a fraction $f_{\rm d,SN} = 50\%$ of these metals are in the form of dust and that the rest is in the gas phase. Because this parameter is very poorly constrained, we chose a value that falls in the middle of the range explored by \citet{Popping2017}, who found that the value of $f_{\rm d,SN}$ mostly affects the dust content of low-mass, low-metallicity galaxies.
The high-velocity shocks produced by the SN explosion will also partially destroy the dust already present near the SN site. The mass $\Delta M_{\rm dest,SN}$ of the dust destroyed in these events is related to the mass $M_{\rm s,100}$ of gas shocked at above $100\,\mbox{km}\,\mbox{s}^{-1}$ via
\begin{equation}
  \label{eq:dMdustSN}
  \Delta M_{\rm dest,SN} = 0.3 \frac{M_{\rm s,100}}{M_{\rm gas}} M_{\rm d},
\end{equation}
where $M_{\rm gas}$ is the local gas mass and $M_{\rm d}$ is the local dust mass; meaning that 30\% of the gas mass in the shocked gas is destroyed. We estimated the shocked gas mass by taking the Sedov solution in a homogeneous medium following \citet{McKee1989}:
\begin{equation}
  \label{eq:Ms100}
  M_{\rm s,100} \simeq \frac{E_{\rm SN}}{0.736\ \left(100\,\mbox{km}\,\mbox{s}^{-1}\right)^2} \simeq 6800\, E_{\rm SN,51}\,\Msun.
\end{equation}

We modelled the dust destruction via thermal sputtering following \citet{Draine1979}, with a destruction timescale given by \citep[][chapter 25]{Draine2011}:
\begin{equation}
  \label{eq:tsputtering}
  t_{\rm sput} \simeq 0.1 \left(\frac{a_{\rm g}}{0.1\,\mu\mbox{m}}\right) \left(\frac{n_{\rm gas}}{\mbox{H}\,\mbox{cm}^{-3}}\right)^{-1} \left(1 + \left(\frac{T}{10^6\,\mbox{K}}\right)^{-3}\right)\,\mbox{Myr}.
\end{equation}
In parallel, the dust content can grow in mass via accretion of metals from the gas phase. We estimate the competition between dust growth and destruction using an approach similar to that of \citet[][albeit simplified]{Dwek1998} or \citet{Novak2012}:
\begin{equation}
  \label{eq:dustmassgrowth}
  \dot{M}_{\rm growth} = \left(1 - \frac{M_{\rm d}}{M_{Z}}\right) \frac{M_{\rm d}}{t_{\rm growth}} - \frac{M_{\rm d}}{t_{\rm sput}},
\end{equation}
where $M_Z$ the local metal mass in both the dust and gas phases, and $t_{\rm growth}$ is the dust growth timescale
\begin{equation}
  \label{eq:tgrowth}
  t_{\rm growth} = 100 \frac{1}{\alpha(T)} \left(\frac{a_{\rm g}}{0.1\,\mu\mbox{m}}\right)^{-2} \left(\frac{n_{\rm gas}}{\mbox{H}\,\mbox{cm}^{-3}}\right)^{-1} \left( \frac{T}{20\,\mbox{K}} \right)^{-\frac{1}{2}} \,\mbox{Myr},
\end{equation}
where $\alpha(T)$ is a the sticking coefficient of metals in the gas phase onto dust. The details of the impact of the choice of the sticking coefficient $\alpha$ are discussed in {Dubois et al. (in prep)}. Here, we used the results from laboratory experiments by \citet{Chaabouni2012}:
\begin{equation}
  \label{eq:stickingcoeff}
  \alpha(T) = 0.95\frac{1 + \beta\frac{T}{T_0}}{\left(1 + \frac{T}{T_0}\right)^\beta},
\end{equation}
with $T_0 = 56\,\mbox{K}$ and $\beta = 2.22$.
Overall, at temperatures below $T \simeq 3\times 10^4\,\mbox{K}$, dust growth happens faster than destruction via thermal sputtering.

\begin{figure*}  
  \centering
  \includegraphics[width=\linewidth]{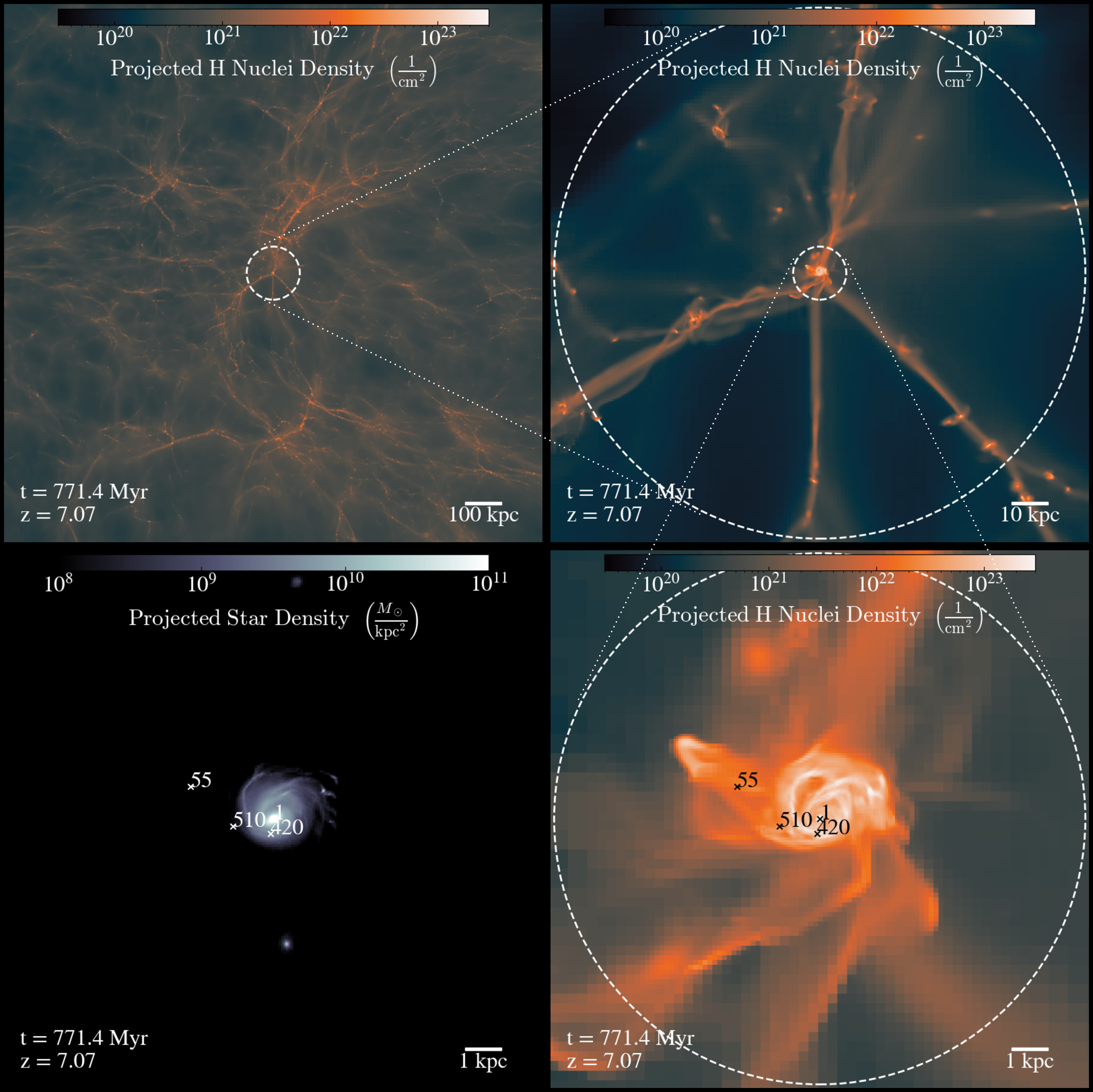}
  \caption{Successive zooms on one of the most massive galaxies in the high-resolution region at $z\sim 7$. From the top left clockwise, the first three panels show the hydrogen column density in a region of dimension $1.5\,\mbox{Mpc}$, $150\,\mbox{kpc}$ and $15\,\mbox{kpc}$ on a side, respectively. The bottom left panel shows the stellar density distribution in the same region as the bottom right panel. The numbered crosses mark the position of the corresponding BHs.}
  \label{fig:zoom_z7}
\end{figure*}

We note here that the dust model is not coupled to the RHD on-the-fly but rather used for post-processing, for instance to assess the UV extinction: Going further would require a significantly more involved dust model \citep[see e.g.][]{Glatzle2019}. To get an estimate of the impact of the UV extinction due to dust, we cast 192 rays from the centre of each galaxy and integrated the dust column density out to the virial radius of the host halo. We then converted this column density to an optical depth using the dust extinction law fits from \citet{Gnedin2008} for the Large Magellanic Cloud. This parametrization assumes that the dust extinction is proportional to the neutral hydrogen column density (their Eq.~5): We rescaled the neutral column density by the dust-to-gas ratio measured in our simulation along each line of sight to measure the dust optical depth.


\begin{figure}
  \centering
  \includegraphics[width=\linewidth]{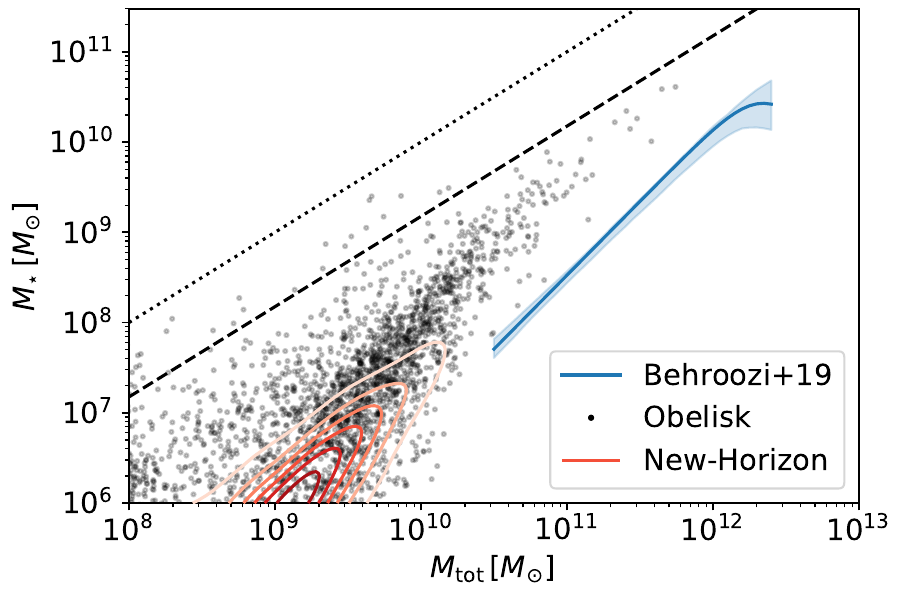}
  \caption{Stellar-to-halo mass relation at $z=6$ in \Obelisk (black dots), compared to the model of \citet[][in blue]{Behroozi2019}, and to the \newh simulation \citet{Dubois2020} (in red) at the same redshift but for an average environment. The dashed and dotted lines indicate the 1:1 relation and the universal baryon fraction, respectively. For haloes with $\Mvir > 10^{9.5}\,\Msun$, the stellar mass in \Obelisk exceeds that of \newh, highlighting the role of the overdensity.}
  \label{fig:SMHM}
\end{figure}

\section{Galaxies and BH populations in \Obelisk}
\label{sec:populations}
In this first paper based on the \Obelisk simulation, we focus on one of the main goals of the project: establishing the respective role of galaxies and accreting BHs in reionizing the large-scale environment of a protocluster.
In this section, we begin by presenting the global properties of both populations; their contribution to reionization will be discussed in the next section.

We illustrate in Fig.~\ref{fig:zoom_z7} the hierarchy of scales captured in the \Obelisk simulation by zooming on one of the most massive galaxies at $z \sim 7.1$, with stellar mass $\Mstar = 2.3\times 10^{10}\,\Msun$. Clockwise, the first three panels show the hydrogen column density \NH at large intergalactic scales ($\sim 1.5\,\mbox{Mpc}$), at the halo scale ($\sim 150\,\mbox{kpc}$) where the large-scale filaments connect to the galaxy, and at galactic scales ($\sim 15\,\mbox{kpc}$) where the galactic disc is roughly face on.
The bottom left panel shows the distribution of stars in the galaxy for the same projection as in the bottom right panel. Even for a galaxy as compact as this one (the effective radius is of order $\lesssim 200\,\mbox{pc}$ and 90\% of the stellar mass is contained within $\sim 850\,\mbox{pc}$ from the centre), we start to capture some of the structure of the disc (e.g. the spiral arms, and a hint of a bar structure).
The numbered crosses in the two lower panels mark the position of each BH, independently of their mass. While there is a massive ($\Mbh \simeq 1.3\times 10^{5}\,\Msun$) BH exactly at the centre of the main galaxy, we can see three other BHs in the image: These are all lower-mass BHs ($\Mbh \simeq 3-4\times 10^4\,\Msun$) and have been brought to the galaxy by previous mergers over the course of its assembly.

We first focus on the galaxies identified in the high-resolution region. Fig.~\ref{fig:SMHM} shows the stellar-to-halo mass relation at $z=6$ for \Obelisk (black dots) compared to estimates from \citet[][blue line]{Behroozi2019}. The red contours show the results from the \newh simulation \citet{Dubois2020}, which features the same subgrid models as \Obelisk but focusing on an average environment. There is no overlap between the models from \citet{Behroozi2019} and \newh, but the extrapolation between the two seems reasonably consistent. By comparison, at high halo masses ($\Mvir > 10^{9.5}\,\Msun$), the galaxies in \Obelisk are more massive than in \newh: Since both simulations share their subgrid models, this is indicative of the effect of the environment.

\subsection{Galaxy populations}
\label{sec:galaxy-populations}

Observations at lower redshift suggest that overdensities are not only comparatively richer in galaxies than the field, but also that the shape of the mass function may depend on the environment \citep[e.g.][]{Davidzon2016, Tomczak2017, Papovich2018}
At a lower redshift, \citet{Shimakawa2018} compare the mass function of galaxies protoclusters to the results for field galaxies in COSMOS \citep{Davidzon2017} at $z \simeq 2-2.5$ and found that at a fixed stellar mass, protoclusters have around 10 times more galaxies.

\begin{figure}
  \centering
  \includegraphics[width=\linewidth]{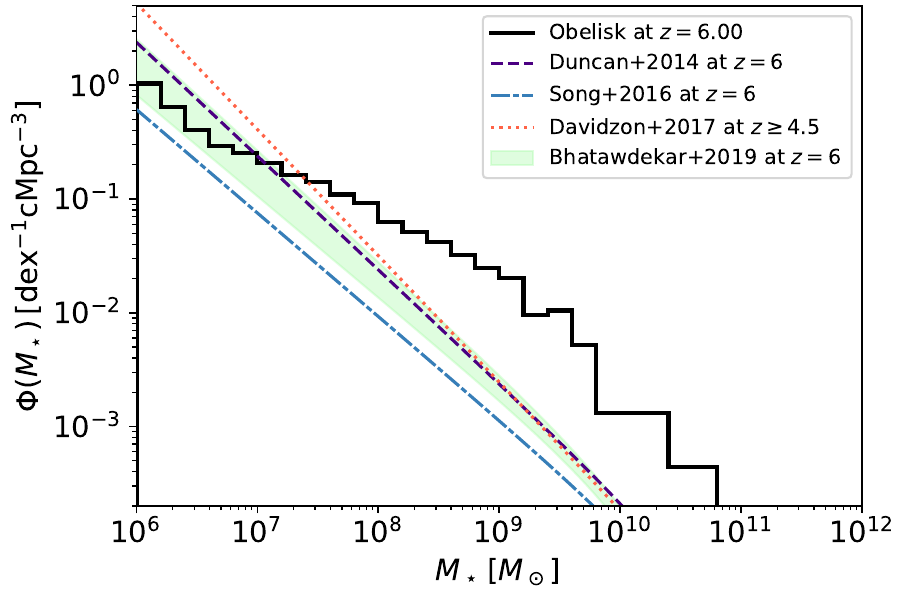}\\
  \includegraphics[width=\linewidth]{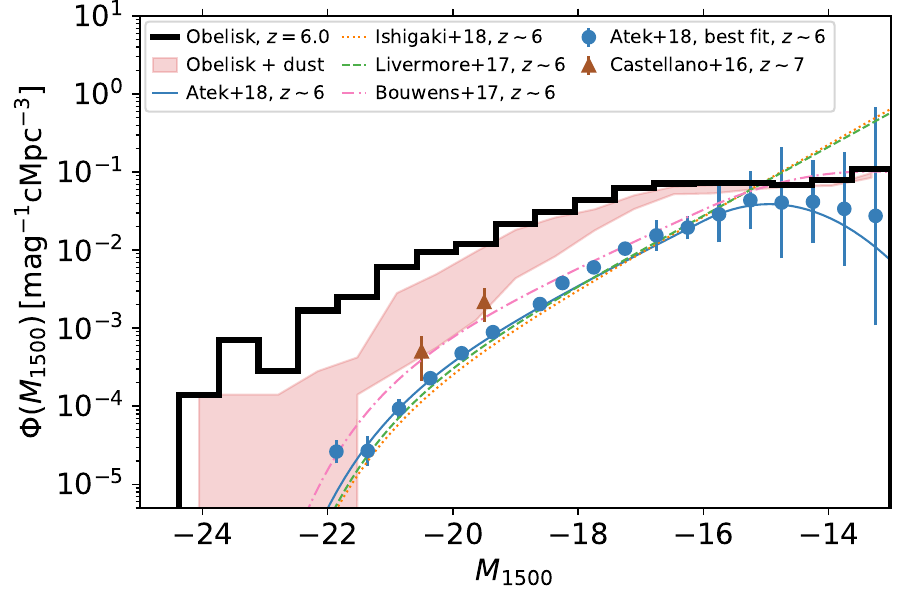}
  \caption{Galaxy mass and luminosity function in \Obelisk. \emph{Top}: Galaxy stellar mass function in our high-resolution region at $z=6$ (thick black line) compared to the observational determination of \citet[dashed purple line]{Duncan2014}, \citet[dash-dotted blue line]{Song2016}, \citet[red dotted line]{Davidzon2017} and \citet[green area]{Bhatawdekar2019}. \emph{Bottom}: Corresponding intrinsic UV luminosity function (thick black line) and including an estimate for the expected dust attenuation based on the dust present in the simulation (red area). We compare our results to the UV luminosity functions for field galaxies found by \citet{Bowler2015, Bouwens2017,Livermore2017,Atek2018,Ishigaki2018} and the BDF field from \citet{Castellano2016}. Details of the legend are given in the text.}
  \label{fig:galaxypops}
\end{figure}

We present in the upper panel of Fig.~\ref{fig:galaxypops} the galaxy stellar mass function at $z\sim 6$, as a thick solid line, and compare our results to average mass functions derived by \citet[dashed purple line]{Duncan2014} and \citet[dash-dotted blue line]{Song2016} based on CANDELS data, \citet[dotted red line]{Davidzon2017} based on the COSMOS survey, and \citet[green area]{Bhatawdekar2019} based on the Hubble Frontier Fields.
As expected, the shape of the mass function is qualitatively different in our simulation, especially at the high-mass end. Around $\Mstar \sim 10^{9-9.25}\,\Msun$, we find a number density around $\Phi \sim 10^{-2}\,\mbox{dex}^{-1}\,h^3\,\mbox{cMpc}^{-3}$. By comparison, \citet{Song2016} find $\Phi \sim 7\times 10^{-4}\,\mbox{dex}^{-1}\,h^3\,\mbox{cMpc}^{-3}$ and \citet{Bhatawdekar2019} find $\Phi \sim 10^{-3}\,\mbox{dex}^{-1}\,h^3\,\mbox{cMpc}^{-3}$, around an order of magnitude below \Obelisk.
At the very low-mass end ($\Mstar \lesssim 10^7\,\Msun$), the number density of galaxies in \Obelisk{} recovers a reasonable agreement with observations (keeping in mind however that, in this mass regime, the assumed morphology of galaxies can change the estimation of the mass function derived by observations by $\sim 0.5\,\mbox{dex}$, \citealp{Bhatawdekar2019}).

The excess of massive galaxies compared to the fields can also be seen in the UV luminosity function (LF). We measure the UV luminosities of our galaxies by assigning a UV luminosity to each star particle in the simulation based on its age, mass and metallicity following the \bpass v2.2.1 SED \citep{Eldridge2017, Stanway2018}, and then summing the luminosity of all star particles associated with a galaxy.
We show the UV LF measured at $z\sim 6$ in the \Obelisk volume (thick solid line) in the lower panel of Fig.~\ref{fig:galaxypops}, compared to the luminosity functions derived from observations of $z\sim 6$ lensed galaxies behind clusters by \citet{Bouwens2017, Livermore2017, Atek2018, Ishigaki2018}. For clarity, we show the best fit LF resulting from these works and only the actual data points from \citet{Atek2018}. We show the data from \citet{Bowler2015} at the bright end, and the determination of the LF by \citet{Castellano2016} for the marginally overdense \emph{Bremer Deep Field} \citep[BDF,][]{Lehnert2003}.

Not only do we expect more galaxies in our simulation (due to the fact that we probe a more biased regions), but we also expect galaxies to be, on average, more evolved at a given redshift compared to the field~\citep[e.g.][]{Overzier2016}. In a $\Lambda$CDM universe, it is expected that the densest structures collapse first, and, due to the so-called halo assembly bias~\citep[e.g.][]{Sheth2004,Harker2006,Borzyszkowski2017,Musso2018}, that galaxies in denser regions are older and more massive compared to the regions of average density.
This is exactly the case here: At $z=6$, galaxies in \Obelisk reach masses in excess of $10^{9-10}\,\Msun$ and UV magnitude brighter than $\MUV \lesssim -22$, and are therefore likely to be significantly enriched in dust.
We repeat the experiment by casting rays from outside of the half-mass radius of each galaxy, to account for stars outside of the centre of the galaxy that might be less attenuated.
This results in the red shaded area: The brighter galaxies are more affected than their faint counterparts. Yet, at all magnitudes, the \Obelisk volume contains more galaxies than the field.
Using simulations based on constrained initial conditions, \citet{Yajima2015} have found a similar increase in the number density of galaxies compared to their simulation of an average patch of the Universe.

\begin{figure}
  \centering
  \includegraphics[width=\linewidth]{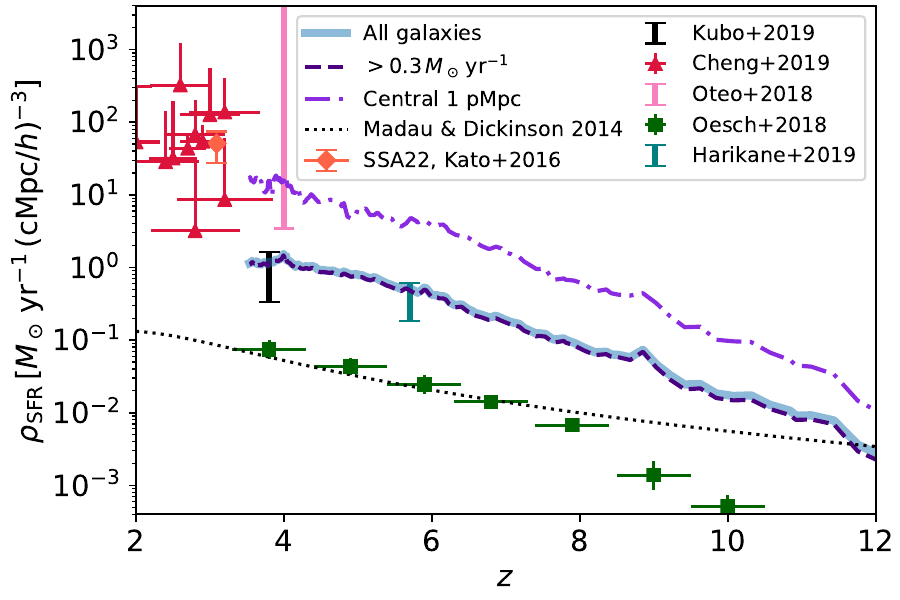}
  \caption{Cosmic SFRD in the \Obelisk volume for all galaxies (in light blue), only for galaxies forming stars faster than $0.3\,\Msun\,\mbox{yr}^{-1}$ (dashed dark purple line), and for all galaxies within 1 Mpc of the most massive galaxy (dash-dotted light purple line). The simulation should be compared to the protocluster observations of \citet[green diamond]{Kato2016}, \citet[black error bar]{Kubo2019}, \citet[red triangles]{Cheng2019} and \citet[teal error bar]{Harikane2019}. The model of \citet[dotted black line]{Madau2014} and the observations of \citet[green squares]{Oesch2018} are only here to guide the eye.}
  \label{fig:rhoSFR}
\end{figure}

We then compare the evolution of the total SFR density $\rho_{\rm SFR}$ measured in the \Obelisk volume to observations both in the field and in high-redshift protoclusters. In Fig.~\ref{fig:rhoSFR}, we show the total $\rho_{\rm SFR}$ as a thick, light blue line. We additionally compute the SFR density for only the galaxies with a SFR higher than $\dot{\Mstar} > 0.3\,\Msun\,\mbox{yr}^{-1}$ (corresponding to $\MUV \simeq -17$, dashed dark purple line) and for galaxies within 1 physical Mpc from the most massive halo (dash-dotted light purple line) as a proxy for the central region of our protocluster. For comparison, we show the best fit cosmic SFR density from \citet[][dotted black line]{Madau2014} and the observational constraints of \citet{Oesch2018} as green squares including previous determinations of the cosmic SFR density from \citet{Oesch2013,Oesch2014,Bouwens2016}.
The other data points are a very heterogeneous compilation of high-redshift protoclusters and overdensities taken from \citet{Kato2016, Harikane2019, Kubo2019} and the compilation of \citet{Cheng2019}. The enhancement of the SFR density measured in \Obelisk compared to the field is in broad agreement with these observations.
For instance, \citet{Harikane2019} found $\rho_{\rm SFR} \simeq 0.32\,\Msun\, \mbox{yr}^{-1} \mbox{cMpc}^{-3}$ at $z \sim 5.7$ for a sample including both \lya emitters (LAEs) and sub-millimetre galaxies, assuming an overdensity radius of $10\,\mbox{cMpc}$. This is very comparable to what we find at $z \sim 6$. At $z\sim 4$, \citet{Oteo2018} detected an overdensity that could correspond to a protocluster core, with a measured SFR around $\sim 6500\,\Msun\,\mbox{yr}^{-1}$ in a projected area of $260\times 310\,\mbox{kpc}^2$ (physical). The total structure, which might extends over more than $2.3  \,\mbox{Mpc}^2$ would have a total SFR of $14\,400\,\Msun\,\mbox{yr}^{-1}$. We bracket these two values in Fig.~\ref{fig:rhoSFR} as the pink error bar.

Similarly, \citet{Kubo2019} quote an average total SFR of $\sim 2.1\times 10^3\,\Msun\,\mbox{yr}^{-1}$ for their $z\sim 3.8$ candidate protoclusters. The upper and lower limits of the error bar in Fig.~\ref{fig:rhoSFR} indicate the resulting SFR density assuming that all the star formation happens in the central (physical) Mpc or in a larger 8 arcmin region, equivalent to $3.4\,\mbox{Mpc}$ (physical) in diameter.
Finally, at a slightly lower redshift, \citet{Kato2016} report a total SFR of $\sim 4.7\times 10^3\,\Msun\,\mbox{yr}^{-1}$ for the concentration of dusty star-forming galaxies in the SSA22 field, corresponding to a SFR density of $\sim 50\,\Msun\, \mbox{yr}^{-1} \mbox{cMpc}^{-3}$ assuming that the protocluster size is around $1\,\mbox{Mpc}$ (physical). While \Obelisk has not reached $z \simeq 3.1$, it seems that extrapolating the trend of the central SFR density would lead to a reasonable (qualitative) agreement with the SSA22 value.

\begin{figure}[!h]
  \centering
  \includegraphics[width=\linewidth]{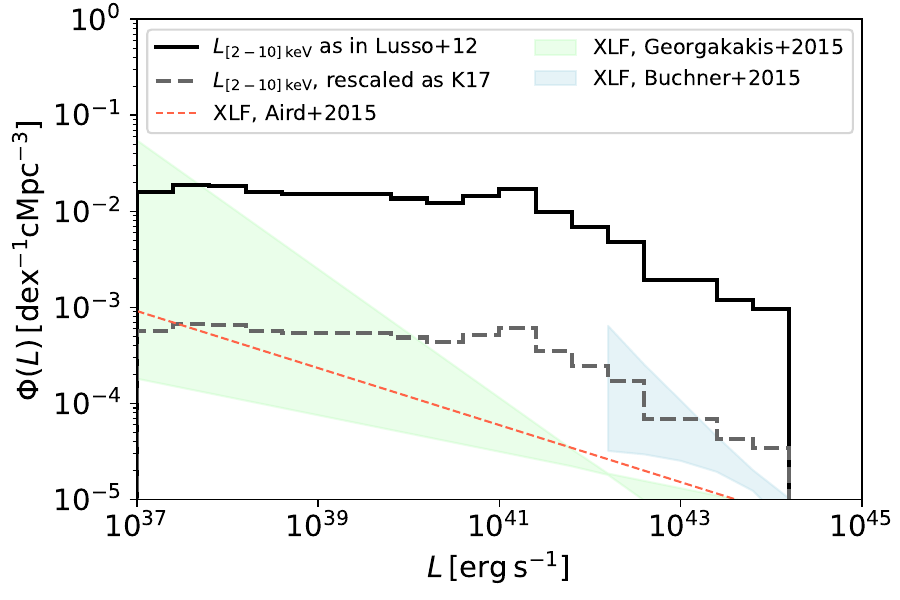}\\[-.5ex]
  \includegraphics[width=\linewidth]{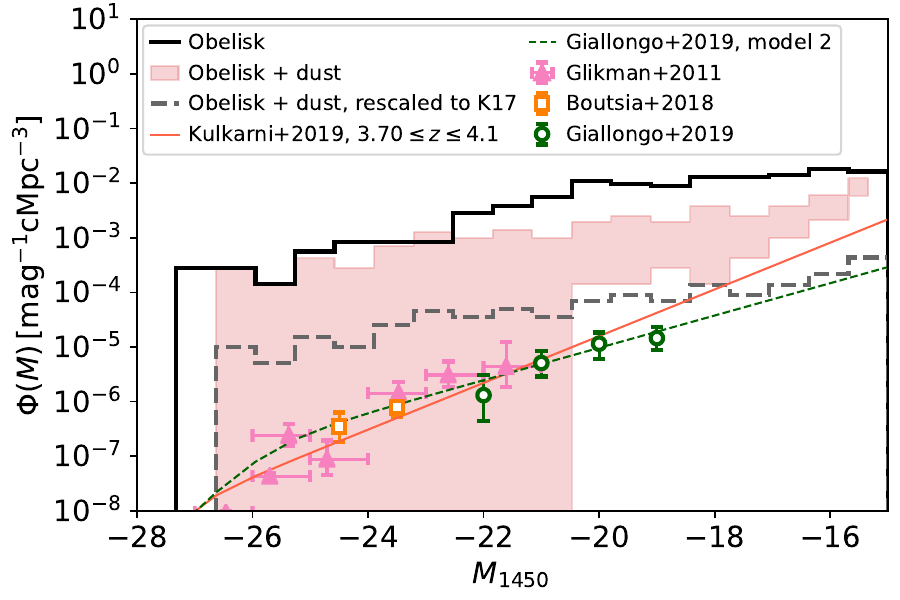}
  \caption{AGN luminosity function. \emph{Top}: AGN X-ray luminosity functions at $z \sim 4$ from the simulation (solid black line) and rescaled to the field (dashed grey line) using the AGN excess found by \citet{Krishnan2017} at $z\simeq 1.6$. We compare our bolometric LF to the fit of \citet[blue line]{Hopkins2007}, our X-ray LF to the results of \citet[red area]{Buchner2015}, \citet[thin dashed line]{Aird2015}, and \citet[green area]{Georgakakis2015}.
    \emph{Bottom}: AGN UV LF at $z\sim 4$ without taking any obscuration into account (solid black line) and using the dust present in the simulation (red area, see text for details), compared to the data of \citet[pink triangles]{Glikman2011}, \citet[orange squares]{Boutsia2018} and \citet[green circles]{Giallongo2019} and to UV LF fits from of \citet[dashed green line]{Giallongo2019} and \citet[red line]{Kulkarni2019}. We rescale again our AGN UV LF with a lower limit of the dust attenuation (dashed grey line) following the excess found by \citet{Krishnan2017}.
  }
  \label{fig:bh-pop}
\end{figure}

\subsection{BH populations}
\label{sec:bh-populations}

Moving on to the BH population of \Obelisk, we show in the upper panel of Fig.~\ref{fig:bh-pop} the AGN luminosity function in the simulation, with the bolometric luminosity function as a solid black line and the hard X-ray luminosity function (XLF) as a dashed red line. We estimate the X-ray luminosity in the $[2-10]\,\mbox{keV}$ band using the bolometric correction from \citet{Lusso2012}. 

As is the case for the galaxy UV luminosity function, most observational estimates focus on the field, and there are no high-redshift samples in overdense regions to compare our results to; the only estimate of the AGN luminosity function in a protocluster \citep{Krishnan2017} is at $z=1.62$. We discuss first the comparison with the field luminosity function, and then discuss the effect of the overdensity on the comparison. 

As expected, our estimated XLF is significantly above the observed XLF in the field \citep[e.g.][]{Ueda2014,Aird2015,Buchner2015,Georgakakis2015,Miyaji2015,Vito2016,Vito2018}. We only show a sample of these observational determinations in the figure for readability. For instance, \citet{Buchner2015} infer $\Phi(L_X=10^{43}\,\mbox{erg}\,\mbox{s}^{-1}) \simeq 4\times 10^{-5} - 2\times 10^{-4}\,\mbox{dex}^{-1}\,h^3\,\mbox{cMpc}^{-3}$ at $4 \leq z \leq 7$, while we find a value between 5 and 25 times higher.  In a protocluster at $z=1.62$ \cite{Krishnan2017} find that the XLF is higher than in the field by a factor of $\sim 28$ at $L_X=10^{43}-10^{44}$, compatible with our result. If we rescale our estimated XLF by dividing the AGN number density by a constant factor of $28$ (dashed grey line), we find a reasonable agreement between the simulation and the observed XLF from \citet{Buchner2015}.

In the lower panel of Fig.~\ref{fig:bh-pop}, we repeat the same exercise for the UV luminosity of the AGN population in \Obelisk, compared to a sample of observations from \citet{Glikman2011}, \citet{Boutsia2018}, and \citet{Giallongo2019}, and to the fits by \citet{Giallongo2019} and \citet{Kulkarni2019}. Here, the AGN overdensity is even more pronounced: This is in part because we have included all accreting BHs from the simulation, and a significant fraction of these are thought to be obscured \citep[e.g.][]{Vito2018}. Obscuration affects somewhat the XLF, via a correction for Compton-thick AGN, although this is mitigated in high-redshift observations because redshift shifts the restframe band to higher energies, and strongly the UVLF.  \cite{Trebitsch2019} using a zoom simulation of a high-redshift galaxy show that UV observations capture only about 3\% of the accreting AGN. We apply a correction for obscuration, similar to what we have done in Fig.~\ref{fig:galaxypops} to allow for a fairer comparison, which, however, still does not account for the \Obelisk region being an overdensity, which as noted above is an additional factor $\sim 28$ in X-ray. The red area on the plot shows the effect of dust attenuation on the UVLF, based on the dust column density measured within $10\,\mbox{kpc}$ from each BH, with the upper limit taking only into account dust beyond $100\,\mbox{pc}$ of each BH (mimicking dust attenuation coming from the transfer through the ISM and CGM). We further rescale this upper limit on the UVLF using the factor of $\sim 28$ from the X-ray (dashed grey line).

\begin{figure}
  \centering
  \includegraphics[width=\linewidth]{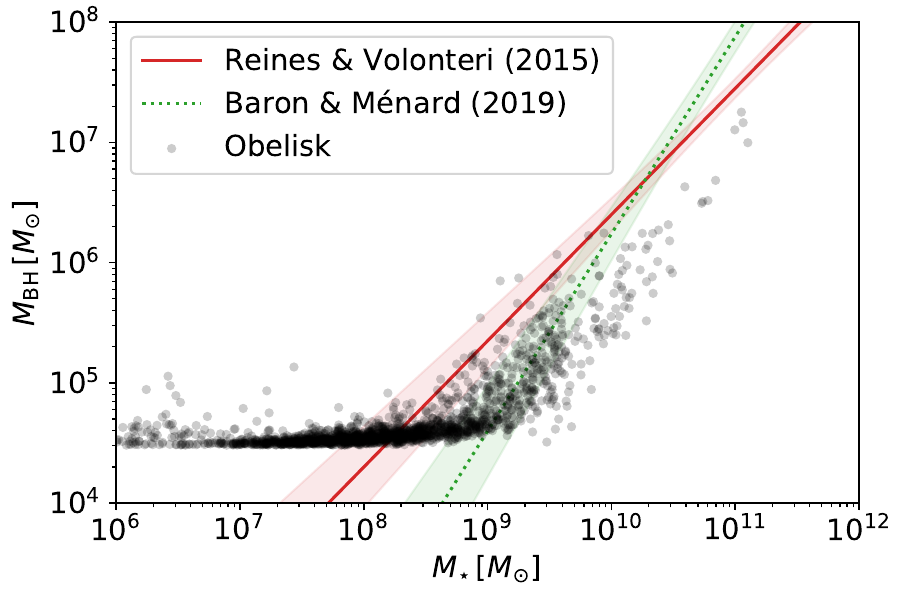}
  \caption{BH mass vs. stellar mass at $z=4$ in \Obelisk (black dots) compared to the observational constraints from \citet{Reines2015} and \citet{Baron2019}. The plateau at low $\Mstar$ corresponds to the mass of the BH seeds.}
  \label{fig:mbhmstar}
\end{figure}

While there is, to the best of our knowledge, no observational constraint on the AGN luminosity function in protoclusters at $z>1.6$, there has been evidence for enhanced AGN activity in dense environments at $z\sim2-3$ \citep[e.g.][]{Lehmer2009, Digby-North2010}. An interesting comparison is in SSA22 at $z\sim3$, where there have been studies on both the galaxy and AGN populations. \cite{Lehmer2009} find that the AGN fraction in SSA22 is increased by a factor of 6 with respect to the field, which means that AGN are enhanced more than galaxies in protoclusters. Similarly, \cite{Digby-North2010} and \cite{Krishnan2017} find that the enhancement in AGN is higher than expected simply taking into account the higher number of galaxies, that is, a higher fraction of  galaxies  in protoclusters exhibit AGN activity. This has been suggested to be a byproduct of protoclusters hosting more more massive galaxies  \citep[see e.g.][for a discussion]{Yang2018} rather than of enhanced interactions or more sustained BH growth.

Finally, in Fig.~\ref{fig:mbhmstar} we compare the BH mass versus the galaxy stellar mass at $z=4$ in \Obelisk with the local scaling relations from \citet{Reines2015} and \citet{Baron2019}. Consistent with previous work \citep[e.g.][]{Dubois2015,Bower2017,Habouzit2017,Weinberger2017}, we find that BH growth is inefficient in galaxies with masses below $\Mstar \lesssim 10^{9.5}\,\Msun$, as indicated by the plateau around the BH seed mass. In the high-mass regime, BHs grow rapidly and start to follow scaling relations similar to those observed at $z \sim 0$. We leave the detailed study of the BH population and its growth history for a future work.

In the context of the $z \gtrsim 6$ Universe, we note that we do not find in \Obelisk any BH with mass of order $\Mbh \gtrsim 10^9\,\Msun$, the expected mass range of BHs powering the brightest quasars observed in the reionization era \citep[e.g.][for recent results]{Mazzucchelli2017b,Banados2018,Reed2019}. This is entirely expected  because these are rare objects with a number density of order $\sim 10^{-9} \,\mbox{cMpc}^{-3}$ \citep{2019ApJ...884...30W} and can therefore only be modelled in simulations representing a much larger volume than \hagn, not even \textsc{BlueTides} simulates a large enough volume \citep{DiMatteo2017}. By construction, as the \Obelisk region was selected as the most massive region in the \hagn box at $z \sim 2$, it corresponds to a number density of order $\sim 10^{-6}\,h^3\,\mbox{cMpc}^{-3}$, we do not expect our volume to contain any bright $z \gtrsim 6$ quasar. Regardless, we note that \Obelisk contains a large number of AGN that can in principle act as sources of reionization, as we explore in Sect.~\ref{sec:sources}.

\section{Reionization of the \Obelisk universe}
\label{sec:reionization}  
\begin{figure*} 
  \centering
  \includegraphics[width=\linewidth]{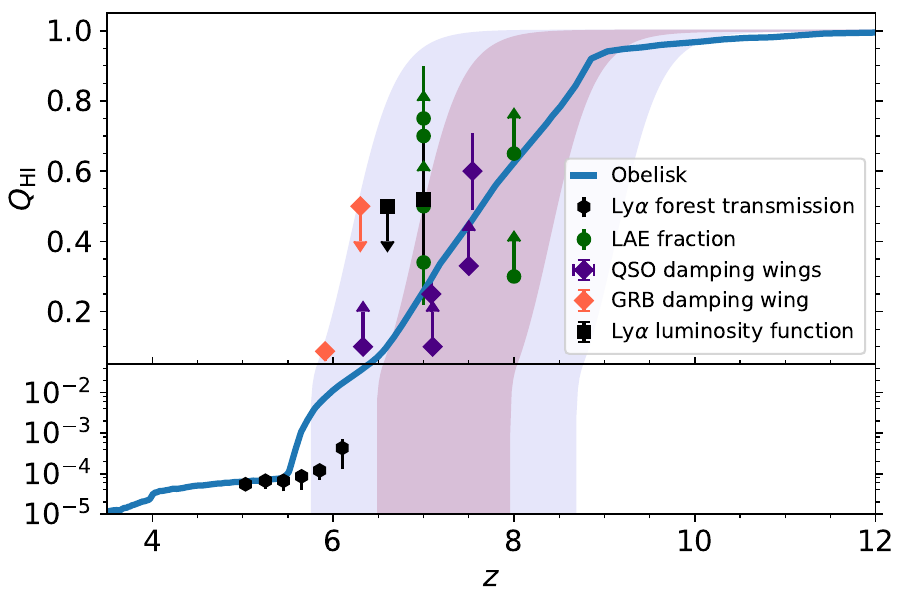}
  \caption{Evolution of the volume-filling fraction of neutral gas in the \Obelisk volume. The shaded area indicate the cosmic microwave background constraints from \emph{Planck}, and the data points show various observational constraints (see text for details). While reionization starts at $z > 12$, the volume is only significantly ionized ($Q_\hi < 90\%$) at $z_{10} \sim 8.68$ and reionization finishes around $z_{99} \sim 5.92$, with a midpoint around $z_{50} = 7.53$.}
  \label{fig:qhi_history}
\end{figure*}

Now that we have established the main properties of the populations of sources (and in particular that the distribution of sources is in line with the expectations for protocluster environments), we can focus on their respective role in the reionization history of the \Obelisk simulation.

\subsection{Hydrogen reionization history}
\label{sec:reionization-history}

\begin{figure*} 
  \centering
  \resizebox*{!}{\dimexpr\textheight-4\baselineskip\relax}{%
    \includegraphics{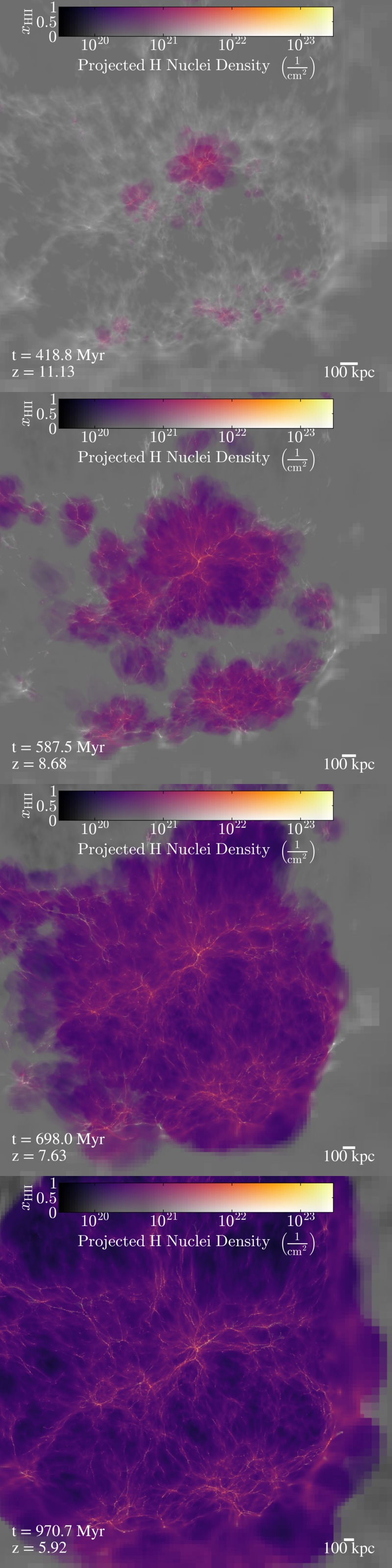}
    \includegraphics{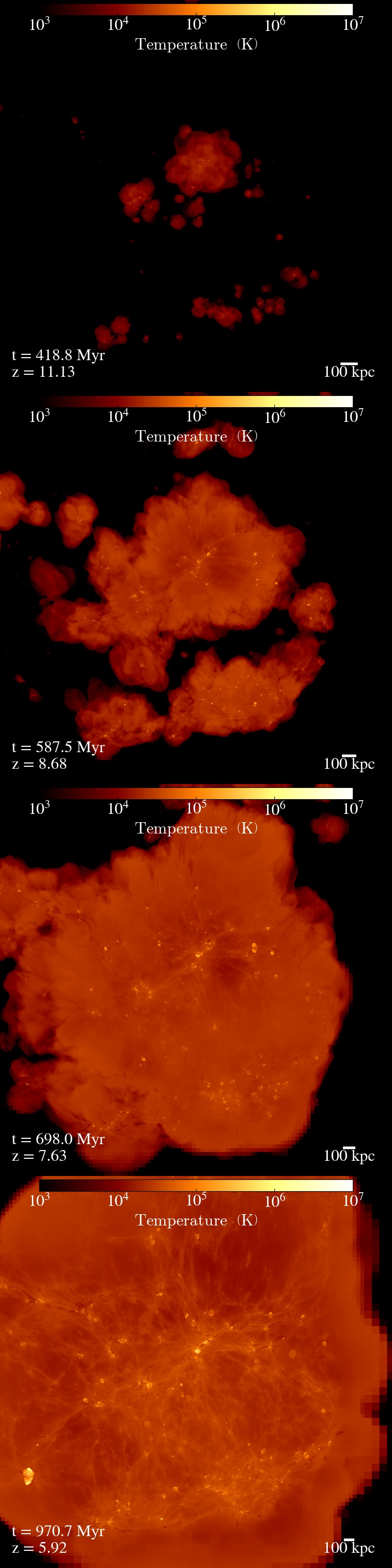}
    \includegraphics{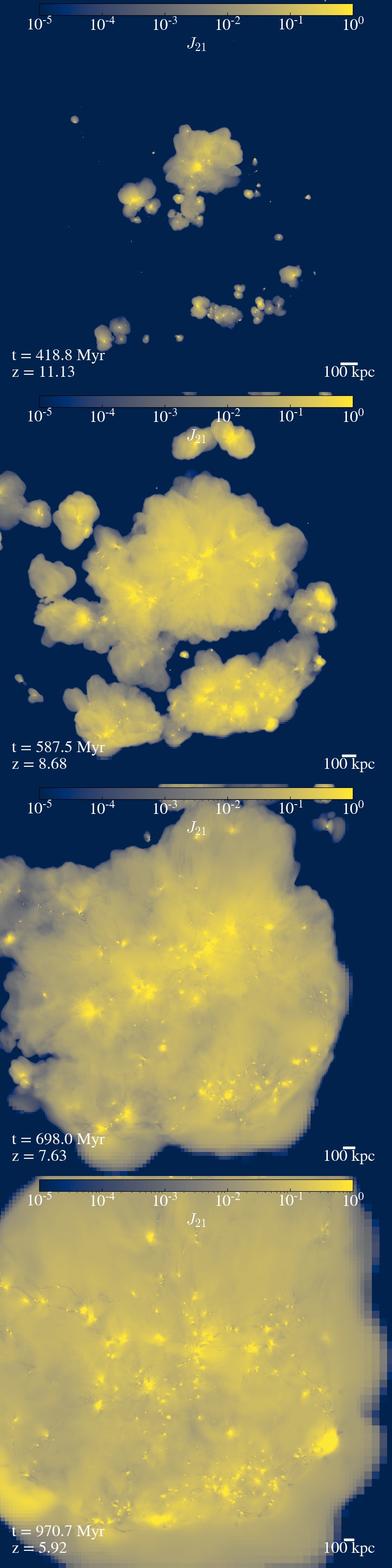}
  }
  \caption{Illustration of the reionization of the volume. The four rows are four different snapshots at $z = 11.13, 8.68, 7.63, 5.92$, corresponding to a ionized volume fraction of the \Obelisk universe of $1\%, 10\%, 50\%$, and $99\%$. The left column shows the gas density (brighter is denser) with the coloured vs. greyscale regions indicating ionized vs. neutral gas. The central column shows the gas temperature, with cold neutral gas in black, warm photoionized gas in dark orange, and hot shocked gas in yellow. The ionized regions can be mapped in the right panel to the local ionizing flux in units of $J_{21} = 10^{-21}\, \mbox{erg}\, \mbox{s}^{-1}\, \mbox{Hz}^{-1}\, \mbox{sr}^{-1}\, \mbox{cm}^{-2}$. All three columns are mass-weighted projections.}
  \label{fig:illustration_reionization}
\end{figure*}

The global volume-weighted neutral fraction of hydrogen $Q_\hi$ in the high-resolution volume is presented as the thick blue line in Fig.~\ref{fig:qhi_history}. The reionization process starts when the first stars are born, and by $z_{50} \simeq 7.63$, half of the volume is reionized. We identify the redshifts at which 1\%, 10\%, 50\%, 90\% and 99\% of the volume is ionized as $z_{01} = 11.13$, $z_{10} = 8.68$, $z_{50} = 7.63$, $z_{90} = 6.58$, and $z_{99} = 5.92$ respectively; corresponding to a reionization duration $\Delta z = z_{99} - z_{10} = 2.8$ ($\Delta t \simeq 385\,\mbox{Myr}$), broadly consistent with the estimates of \citet{Robertson2015}
The dark and light shaded areas in Fig.~\ref{fig:qhi_history} correspond to the $1\sigma$ and $2\sigma$ constraints on the redshift of reionization from the cosmic microwave background measurements of the \emph{Planck} mission \citep{PlanckCollaboration2018}, with a reionization midpoint $z_{\rm re} = 7.67 \pm 0.73$. We also compare the \Obelisk reionization history to a selection of observational constraints: Black hexagons correspond to the measurements of the Lyman-$\alpha$ forest transmission~\citep[\lya forest,][]{Fan2006}, the green circles show constraints on the IGM opacity from the fraction of Lyman-$\alpha$ emitters in Lyman-break galaxy samples \citep[][]{Schenker2014, Ono2012, Pentericci2014, Robertson2013, Tilvi2014}, the purple diamonds show measurements from quasar damping wings by \citet{Mortlock2011, Schroeder2013, Banados2018, Durovcikova2020}, the red diamonds show similar measurements on gamma-ray bursts \citep[GRB,][]{Totani2006, Totani2016}, and the black squares from \citet{Ouchi2010, Ota2008} represent constraints derived from the evolution of the Lyman-$\alpha$ luminosity function. Some of these data points come from the compilations of \citet{Bouwens2015}.
Overall, we find that the simulation agrees with most observations in terms of reionization history, despite the fact that we focus on an overdense region.
Interestingly, the simulation manages to capture residual neutral fraction after reionization is complete at $z < 6$ similar to what is observed. We discuss this point further below.

We illustrate the reionization process on the scale of our high resolution region in Fig.~\ref{fig:illustration_reionization}. The four rows represent four different snapshots of the simulation, at $z = 11.13, 8.68, 7.63, 5.92$, corresponding to a ionized volume fraction of $1\%, 10\%, 50\%$, and $99\%$. Each panel is a projection in a $20 \times 20 \times 2\,h^{-3}\,\mbox{cMpc}^3$.
The first column shows both the gas density and the ionization state of the gas. Brighter regions on the maps are denser, and colourful regions are ionized while grey regions are still neutral. We see that as ionized bubbles grow and expand, the matter collapsed in haloes and voids appear on large scales. While the bottom row corresponds to a $99\%$ ionized volume, we see some still neutral regions remaining on the map: This illustrates the contours of the high-resolution region well.
The second column shows the temperature distribution: Ionized regions reach $T \gtrsim 1-2 \times 10^4\,\mbox{K}$, and we can identify hot bubbles around the knots of the cosmic web that are created by feedback from AGN and supernovae.
The third column presents the ionizing flux in units of $J_{21} = 10^{-21}\, \mbox{erg}\, \mbox{s}^{-1}\, \mbox{Hz}^{-1}\, \mbox{sr}^{-1}\, \mbox{cm}^{-2}$. We can again identify the ionized regions at early time (before overlap) and the contours of the high-resolution region in the last panel. Interestingly, even when the region is completely ionized, we note that the ionizing flux varies by more than two orders of magnitude.

\begin{figure}
  \centering
  \includegraphics[width=.95\linewidth]{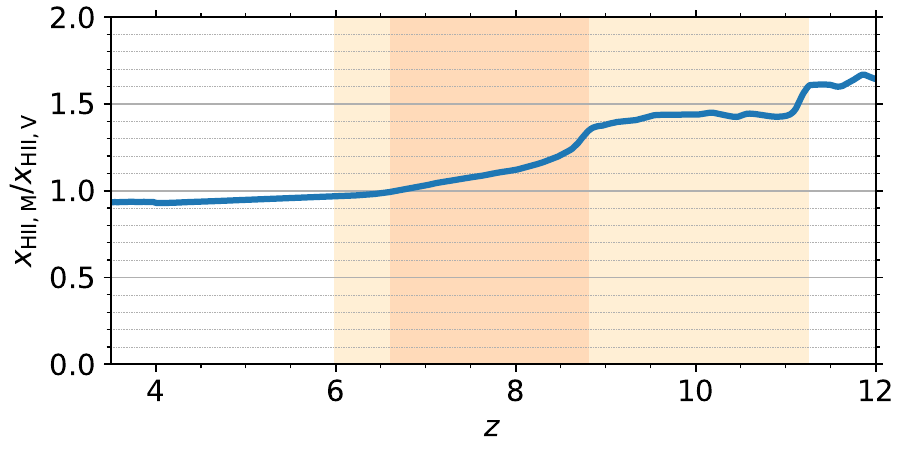}
  \caption{Ratio of the mass-weighted ionized fraction $x_{\hii,\rm M}$ to the volume-weighted ionized fraction $x_{\hii,\rm V}$, showing how reionization happens inside-out: Overdense regions get ionized first by their central sources ($x_{\hii,\rm M} > x_{\hii,\rm V}$), followed by voids. The final ratio is just below 1 because the gas in the densest regions (i.e. haloes and filaments) can recombine, so that $x_{\hii,\rm M} \lesssim x_{\hii,\rm V}$ after reionization is complete. The light (dark) shaded region marks the $z_{01}-z_{99}$ ($z_{10}-z_{90}$) redshift intervals.}
  \label{fig:qhi_ratio}
\end{figure}
This good agreement with observational constraints is not necessarily expected: While overdense regions such as the one we are modelling here are rich in ionizing sources (as shown in the previous section), there is also more gas to reionize compared to a field environment. Earlier studies have found conflicting results: For instance, the simulations of \citet{Ciardi2003} have suggested that protoclusters could actually be completely reionized later than average environments, while the beginning of the reionization process happens earlier. Using a semi-analytical approach, \citet{Kulkarni2011} found that on the contrary, overdense regions are reionized earlier in their model. More recent very large-scale simulations for instance by \citet{Iliev2006,Iliev2014} point towards the same direction: They found a positive correlation between the reionization redshift and the overdensity of the region.
We should note, however, that reionization simulations have tremendously progressed these past few years: For instance, \Obelisk has a mass resolution 100 times better than the \citet{Ciardi2003} simulation, and evolves the radiation field directly coupled with the hydrodynamical evolution. Nevertheless, this suggests that the complex balance between an increased number of both sources and sinks of ionizing photons needs to be studied in more detail.
The detailed comparison between \Obelisk and these earlier works is difficult: The correlation between reionization history and density will depend on how the environment affects the source properties, such as the escape fraction or star formation. In \citet{Ciardi2003}, a constant \fesc is assumed, and no radiative feedback is implemented. By comparison, \citet{Kulkarni2011} and \citet{Iliev2014} both assume that star formation is suppressed in the lowest-mass haloes and incorporate \fesc in their source model. Since these works aim to match the average reionization history, suppressing star formation in low-mass haloes will lead to a more clustered source distribution and could explain the ionization history - density correlation they suggest. In our study, these effects are taken into account self-consistently. However, we did not disentangle the effects of Jeans suppression and varying \fesc or luminosity with environment: we defer the study of the role of environment on the ionizing output of galaxies to a future work.
In \Obelisk, we find that even though our reionization mid-point is fully consistent with constraints on the average reionization redshift, the end of reionization is a bit delayed: $z_{99} = 5.92$. This can be seen again in Fig.~\ref{fig:qhi_history}: The simulation overshoots the data points from \citet{Fan2006}. It is entirely possible that this comes from the details of the simulation setup, but this is still suggestive that the very end of reionization happens later in \Obelisk than in an average environment. By comparison, with a similar simulation setup\footnote{This also comes from the fact that \citet{Rosdahl2018} used the v2.0 of the \bpass library. Using \bpass v2.1.0, they find a significantly delayed reionization {Rosdahl et al. (in prep)}.} (albeit with higher spatial resolution), \citet{Rosdahl2018} finds that the \sphinx volume is 99\% reionized at $z \sim 7$. This is reminiscent for instance of the results of \citet{Aubert2018}, who showed using the \textsc{CoDa I-AMR} simulation that in environments similar to the Local Group, reionization of progenitors of the most massive haloes starts earlier but last significantly longer.

The fact that our residual neutral fraction at $z < 6$ is consistent with the observations might seem at first in contradiction with the results of \citet{Ocvirk2019}. Indeed, with our type of implementation of the `reduced speed of light' approximation, \citet{Ocvirk2019} found that the residual neutral fraction after reionization is complete scales inversely with the adopted speed of light reduction factor $f_c$. This is a consequence of the fact that the photoionization equilibrium of strongly ionized gas can be approximated by $x_\hi \sim \frac{\alpha_{\rm B} \rho_{\element{H}}}{\rho_\gamma \sigma f_c c}$ where $\alpha_{\rm B}$ is the case-B recombination coefficient, $\rho_{\element{H}}$ the total hydrogen density, and $\rho{\gamma}$ the ionizing photon density.
In our case, there are however two mitigating factors to this problem. First, the use of the VSLA has been shown by \citet[Appendix B]{Katz2018} to yield converged results for reduction factors of order $f_{c, \rm max} \sim 0.1-0.4$, and we use $f_{c, \rm max} = 0.2$.
A second difference is that our cosmological setup is very different compared to \citet{Ocvirk2019}: Because we zoom within the \hagn volume, radiation is allowed to leak out of the high-resolution region. The argument in \citet{Ocvirk2019} is that, in the post-overlap universe, the volume is roughly at photoionization equilibrium and all emitted photons contribute to this equilibrium. Here, the situation is different: Photons can either be absorbed within the high-resolution region or escape the volume. This will shift the photoionization equilibrium towards a higher neutral fraction. This also means that the dependence on $f_c$ is reduced: A higher (lower) reduced speed of light would lead to more (fewer) photons escaping the volume by the same redshift, shifting the equilibrium towards a higher (lower) neutral fraction, counterbalancing somewhat the effect found by \citet{Ocvirk2019}.

To explore this picture in which reionization happens first inside-out and then outside-in, we follow the approach of \citet{Iliev2006,Bauer2015} and show in Fig.~\ref{fig:qhi_ratio} the ratio of the ionized mass fraction to the ionized volume fraction, $x_{\hii,\rm M} / x_{\hii,\rm V}$. This ratio is directly related to the (over)density of ionized gas $\delta_\hii$ via\footnote{If we note $M_{\rm box}$ and $V_{\rm box}$ as the mass and volume of the box, we have: $x_{\hii,\rm M}/x_{\hii,\rm V} = x_{\hii,\rm M} M_{\rm box}/V_{\rm box} \times V_{\rm box}/(x_{\hii,\rm V} M_{\rm box})  = (x_{\hii,\rm M} M_{\rm box})/(x_{\hii,\rm V} V_{\rm box}) \times V_{\rm box}/M_{\rm box} = M_\hii / V_\hii \times V_{\rm box}/M_{\rm box} = \rho_\hii / \bar{\rho}$.} $x_{\hii,\rm M} / x_{\hii,\rm V} = \rho_\hii/\bar{\rho} = 1 + \delta_\hii$, with $\rho_\hii$ the density of ionized hydrogen in the volume and $\bar{\rho}$ the universal cosmic hydrogen density. Here, the light (dark) shaded region marks the $z_{01}-z_{99}$ ($z_{10}-z_{90}$) redshift intervals. We see that at early times, the ratio is above unity, indicating that $\delta_\hii > 0$: Early ionized regions typically correspond to overdensities. At later times, this ratio decreases, and by the time the \Obelisk universe is 90\% ionized, ionized regions are typically at the average density. Interestingly, $x_{\hii,\rm M} / x_{\hii,\rm V}$ seems to converge to a value right below unity: This is because the collapsed regions (e.g. haloes) are included in this analysis, and they represent the densest regions where gas can recombine efficiently.

\subsection{Relative contribution of galaxies and black holes}
\label{sec:sources}

\begin{figure}
  \centering
  \includegraphics[width=\linewidth]{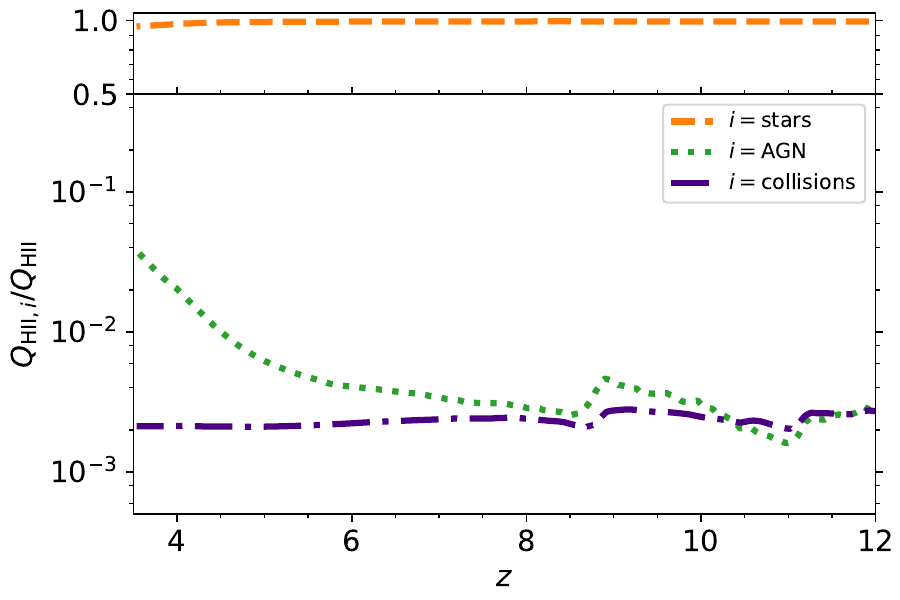}
  \caption{Contributions of various sources to the volume-weighted ionization: At all times, most of the ionized volume is ionized by stellar populations, while AGN only start to be relevant at $z \lesssim 4$.}
  \label{fig:qhii_fractions}
\end{figure}
Using the photon tracer method of \citet{Katz2018}, we can now to relate the source populations described in Sect.~\ref{sec:populations} to the reionization history presented in the previous section.
To this end, we measure what fractions of the ionized volume have been carved out by stellar photons, AGN photons, and collisional ionizations. These fractions are shown in Fig.~\ref{fig:qhii_fractions}. The orange dashed line corresponds to the contribution of photons of stellar origin, and dominates overwhelmingly the contributions of photons produced by AGN (green dotted line) and collisional ionization (purple dash-dotted line). In the rest of this section, we use the same colour and line-style convention for all figures, unless stated otherwise.
This means that the \Obelisk volume is predominantly reionized by stellar populations. We note, however, that at $z \lesssim 5$, the contribution to photo-ionization by AGN becomes more and more important, consistent with the picture that AGN maintain the ionization state of the Universe post-reionization \citep[e.g.][]{Haardt2012, Becker2013, Faucher-Giguere2020}.
Overall, collisional ionizations are completely irrelevant: This is not unexpected as they predominantly occur in the vicinity of galaxies and haloes, which are already not very volume-filling and already largely photo-ionized.

\begin{figure}
  \centering
  \includegraphics[width=\linewidth]{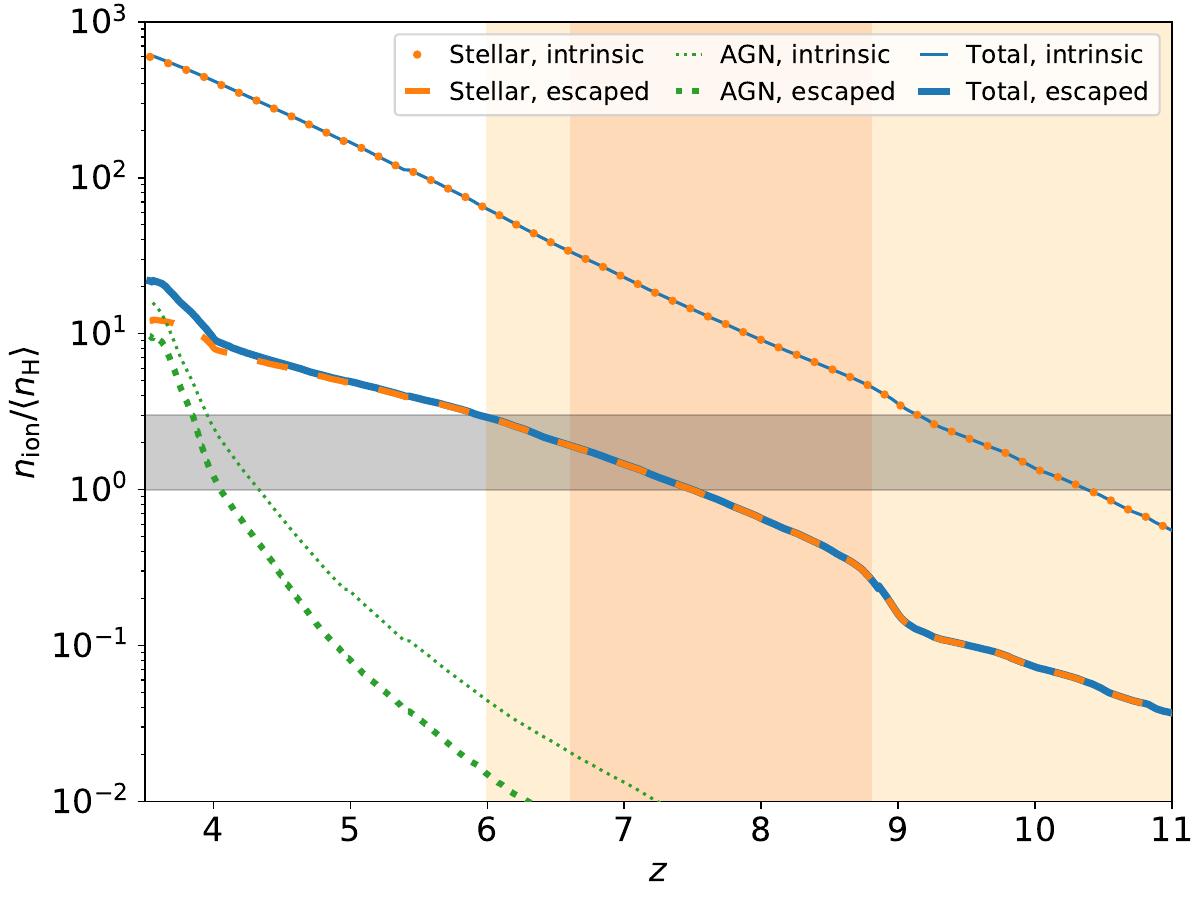}
  \caption{Cumulative ratio of the number of produced (thin lines) and escaped (thick lines) ionizing photon to the average hydrogen density. The solid blue, dashed orange and dotted green lines correspond to the total contribution and that of stellar populations and AGN, respectively. For better readability, we show the intrinsic stellar contribution with dots instead of a line. The background shaded area indicates the $z_{01}-z_{99}$ ($z_{10}-z_{90}$) redshift intervals. The grey area highlights a photon-to-baryon ratio between 1 and 3, necessary to reionize the Universe. Stellar populations alone provide the necessary number of photons by $z \sim 6$.}
  \label{fig:photon_to_baryon}
\end{figure}
We explore this further in Fig.~\ref{fig:photon_to_baryon}, which shows the cumulative photon-to-baryon ratio for different sources. The intrinsic ratio, that is the total number of photons produced divided by the total hydrogen number density, is depicted with thin lines, while the ratio of photons in the IGM over total hydrogen number density is shown with thick lines. Here, we define photons in the IGM (escaped photons) as photons in cells where the gas density is below 180 times the average density at that redshift.
The light (dark) shaded region marks again the $z_{01}-z_{99}$ ($z_{10}-z_{90}$) redshift intervals, and the grey horizontal region highlights a photon-to-baryon ratio between 1 and 3, corresponding to the typical photon budget required to reionize the Universe.
The most striking feature of this figure is that until $z \lesssim 4$, AGN are completely irrelevant to the photon budget of the Universe, despite the fact that we are studying a region that is particularly rich in AGN. Stellar populations alone account for most of the photons, both intrinsically and after transfer through the ISM. Around our reionization midpoint, they have contributed around 1 photon per atom, and by $z_{99}$, their contribution reaches the critical value of around 3 photons per hydrogen atom. By comparison, AGN reach this value only at $z \sim 4$.
We note however that, analogous to Fig.~\ref{fig:qhii_fractions}, the AGN contribution increases quickly at $z \lesssim 4$, but still represents a small fraction of the number of photons in the IGM.

\begin{figure}
  \centering
  \includegraphics[width=\linewidth]{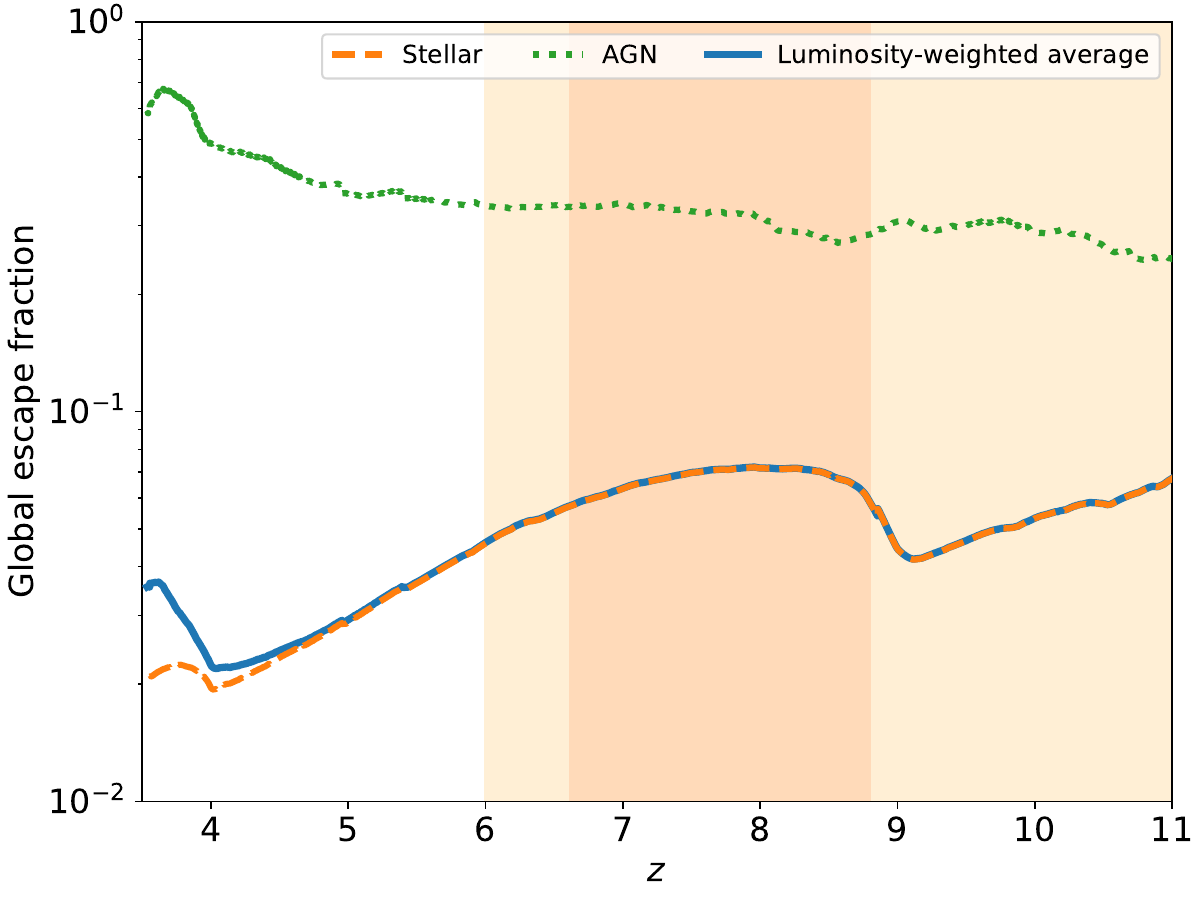}
  \caption{Global escape fraction for different sources with the same colour scheme as in Fig.~\ref{fig:photon_to_baryon}, defined as the total escaped luminosity of a population divided by its intrinsic luminosity. The luminosity-weighted average escape fraction is very close to that of stellar populations at $z \gtrsim 4$, highlighting that they largely dominate the photon budget.}
  \label{fig:global_fesc}
\end{figure}

A second aspect of Fig.~\ref{fig:qhii_fractions} is that on average, the fraction of photons of stellar origin escaping into the IGM is lower than for the photons of AGN origin. Indeed, the ratio of the thick to thin lines correspond to the population-averaged escape fractions, $\langle\fesc\rangle$. We show this global escape fraction in Fig.~\ref{fig:global_fesc} for the different source populations.
We choose to leave the analysis of the variation in \fesc on an object-by-object basis to a forthcoming work, but we can note two main results from this.
Overall, AGN have a very high (population-averaged) escape fraction, of the order of $\langle\fescAGN\rangle \sim 20-50\%$. This is in good agreement with the estimates of \citet{Cristiani2016} at lower redshift, although this does not correspond to the escape fraction of individual AGN in our simulation. We defer a detailed analysis of \fescAGN to a further study.
By comparison, stellar sources exhibit $\langle\fescstar\rangle \lesssim 10\%$, consistent with the values found by other high-resolution reionization simulations \citep[e.g.][]{Kimm2014, Ma2015, Xu2016, Rosdahl2018, Yoo2020}. Here again, we stress that this is a value averaged over the whole population, and that the individual $\fescstar$ can vary by orders of magnitudes between objects and even for a given galaxy over the course of its evolution \citep[e.g.][]{Kimm2014, Wise2014, Paardekooper2015, Trebitsch2017}.
Nevertheless, the population-averaged $\langle\fescstar\rangle$ (or $\langle\fescAGN\rangle$, for that matter) is a useful figure to compare to global reionization models.

\begin{figure}
  \centering
  \includegraphics[width=\linewidth]{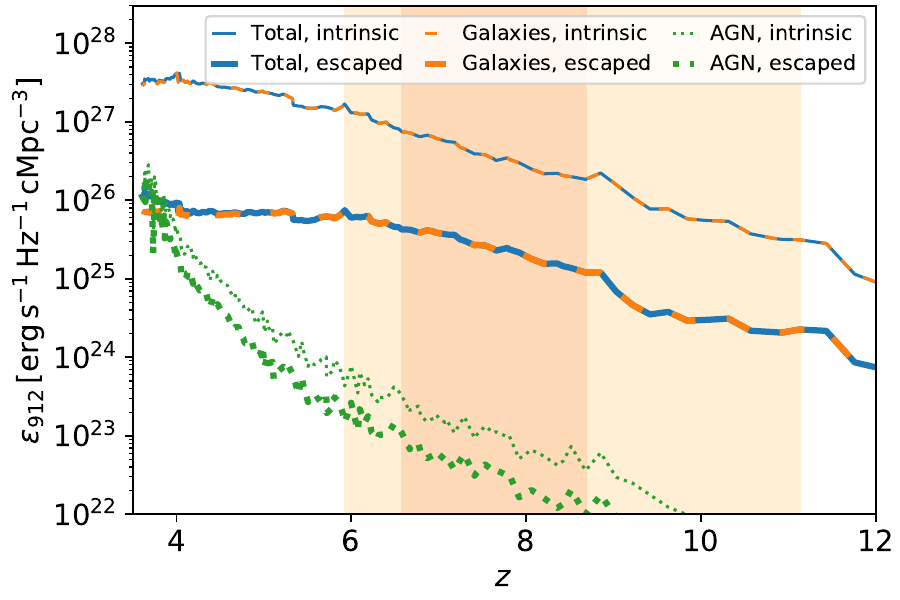}
  \caption{Ionizing emissivity, intrinsic (thin lines) and after transfer in the ISM (thick lines), keeping again the same colour coding as in Fig.~\ref{fig:photon_to_baryon}.}
  \label{fig:global_e912}
\end{figure}

\begin{figure*}
  \centering
  \includegraphics[width=.9\linewidth]{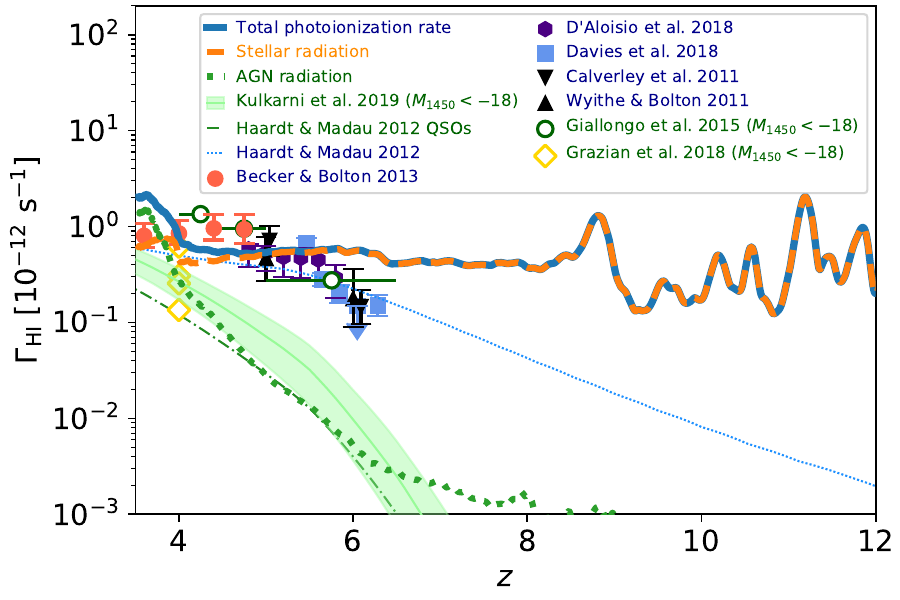}
  \caption{Evolution of the \hi photoionization rate in ionized gas (thick blue line) and the contributions of stellar populations (thick orange dashed line) and AGN (thick green dotted line), compared to the models of \citet{Haardt2012} (thin dotted blue and dash-dotted green lines) as well as the constraints from the homogenized sample of \citet{Kulkarni2019} extending the AGN UV luminosity functon down to $M_{1450} = -18$ and other measurements (see text for details). Blue, orange and green labels correspond to the total photoionization rate, the stellar, and the AGN contribution, respectively. At $z \gtrsim 4$, and in particular during the whole EoR, the \hi photoionization rate is completely dominated by the contribution of stellar populations.}
  \label{fig:PI_hi_history}
\end{figure*}

Combining our estimate for the source populations and their respective escape fractions, we can now directly estimate the emissivity of both types of sources as this is the quantity that typically enters in reionization models. We measure this by summing over all sources (stellar populations or accreting BHs) their ionizing luminosity at $912\,\AA$ using their respective SED (see Sect.~\ref{sec:stellar-radiation} and \ref{sec:bh-radiation}), divided by the total volume of the high-resolution region: This results in the intrinsic emissivity $\epsilon_{912}^{\rm int}$ shown in Fig.~\ref{fig:global_e912} as thin lines with the same colour-coding as previously. We then combine this with our global escape fractions to get the actual emissivity $\epsilon_{912} = \langle\fesc\rangle \epsilon_{912}^{\rm int}$, shown as thick lines in Fig.~\ref{fig:global_e912}.
We observe the same pattern as previously: Galaxies dominate the ionizing photon production, even after transfer through the ISM. The total emissivity $\epsilon_{912}$ is higher by a factor of $\sim 10$ compared to models of \citet{Haardt2012} or \citet{Faucher-Giguere2020}, for instance: This results directly from the fact that our source density is significantly higher than in average environments (see Sect.~\ref{sec:populations}).

Finally, we plot in Fig.~\ref{fig:PI_hi_history} the \hi photoionization rate $\Gamma_\hi$ in ionized gas as a function of cosmic time. The thick lines correspond to the simulation: Once again, the total $\Gamma_\hi$ from all sources is the solid blue line, the contribution $\Gamma_\hi^{\star}$ from stars is illustrated as the dashed orange line, and the contribution from AGN $\Gamma_\hi^{\rm AGN}$ is shown as the dotted green line.
We include the model of \citet{Haardt2012} as a thin dotted blue line, and the contribution from quasars as a thin dash-dotted green line. The green shaded area corresponds to the determination of the contributions of quasars to $\Gamma_\hi^{\rm AGN}$ by \citet{Kulkarni2019} integrating the quasar UV LF down to $M_{1450} = -18$. We also plot various \hi photoionization rates measurements taken from the compilation of \citet{Kulkarni2019}: \citet{Calverley2011} as downward pointing black triangles, \citet{Wyithe2011} as upward pointing black triangles, \citet{Becker2013} as red circles, \citet{DAloisio2018} as purple hexagons, and  \citet{Davies2018} as blue squares. We also show the value of $\Gamma_\hi^{\rm AGN}$ derived by \citealt{Kulkarni2019} based on the \citet{Giallongo2015} AGN luminosity function as empty green circles. The four empty yellow diamonds correspond to the estimates of $\Gamma_\hi^{\rm AGN}$ by \citet{Grazian2018} based on the luminosity functions of \citet{Giallongo2015, Glikman2011, Parsa2018, Akiyama2018}, from top to bottom. These estimates give a measure of the effect of the uncertainty on the faint end of the AGN UV luminosity function on the determination of the contributions of quasars to the \hi photoionization rate.
For most of the reionization era, the simulated \hi photoionization rate remains around $\Gamma_\hi \simeq 5\times 10^{-12}\,\mbox{s}^{-1}$ and is dominated by the contribution of stellar populations. The initial large fluctuations at $z \gtrsim 9$ correspond to the very early stages of reionization: At this epoch, only a small fraction of the volume is ionized (see e.g. Fig.~\ref{fig:qhi_history}), so the value of of $\Gamma_\hi$ will be very sensitive to the stochasticity of both star formation and the subsequent escape of ionizing radiation. Perhaps more interesting, it appears that AGN start to represent a significant contribution to the \hi photoionization rate after $z \lesssim 4$. This is consistent with estimates from e.g. \citet{Kulkarni2019} for the whole AGN population\footnote{We should note that the determination of the faint end of the AGN UV luminosity function has been heavily discussed in the literature recently \citep[e.g.][but see also \citealt{Giallongo2019} for an extended discussion]{Vito2016,Parsa2018}.} (i.e. not only overdense regions), and suggests that even in region where the growth of AGN is favoured, they are not important contributors to the reionization of their large scale environment.

\subsection{Helium reionization}
\label{sec:helium-reionization}

\begin{figure}
  \centering
  \includegraphics[width=\linewidth]{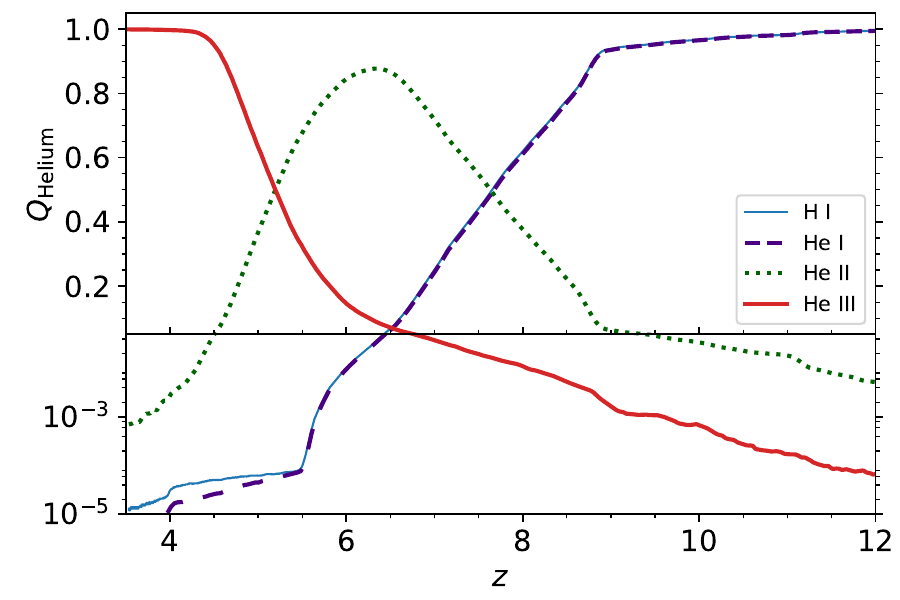}
  \caption{Helium ionization history in the \Obelisk volume. The evolution of the neutral fraction of helium, $Q_\hei$, follows that of the neutral hydrogen $Q_\hi$. \heii reionization is complete by $z \gtrsim 4$: Our region corresponds to one of the growing \heiii bubbles around massive hosts.
  }
  \label{fig:helium}
\end{figure}

Because \Obelisk includes a model for the production of far-UV radiation by AGN, we can follow the ionization state of helium through cosmic time. Indeed, while helium is singly ionized by the same sources that ionize hydrogen, the second ionization of helium can only happen when through photoionization by photons with energies above $54.4\,\mbox{eV}$, which are almost entirely produced by AGN.
We show the \Obelisk helium reionization history in Fig.~\ref{fig:helium}, but leave a more detailed discussion on the properties of \heii-reionization sources to a future paper.
The volume fractions of neutral, singly and doubly ionized helium are shown as a purple dashed line, a green dotted line, and a solid red line, respectively. As expected, the \hei volume fraction follows a trend very similar to that of \hi reionization.
The double reionization of helium starts before \hei single reionization is complete, and finishes fairly early ($z \gtrsim 4$) compared the predicted $z\sim 3$ \heii reionization redshift \citep[e.g.][]{Haardt2012}.

Albeit perhaps surprising, the fact that helium is doubly reionized earlier than for the average universe still results from our choice to model an overdensity.
In the case of \hi reionization, the dominant sources are (faint) galaxies, which are not strongly clustered. In particular, the typical size of an \hii region around galaxies (before overlap) is small compared to the size of our high-resolution region. The situation is very different for \heii reionization, for which the sources are very clustered. The models of \citet{McQuinn2009} or \citet{Dixon2014} suggest that \heiii bubbles around bright sources can extend beyond $R_\heiii \gtrsim 35\,\mbox{Mpc}$ even at $z \sim 6$ (when helium is on average not doubly ionized).
This scale is larger than the size of our high-resolution: In other words, we are probing the expansion of \heiii bubbles around sources in an epoch where the majority of the Universe is still not affected by these bubbles.

\section{Summary and conclusions}
\label{sec:summary}

We have introduced the \Obelisk project: a fully coupled radiation-hydrodynamical cosmological simulation that follows the assembly of a massive protocluster during the first few billion years of its history.
This simulation combines the power of modern cosmological codes to simulate a large overdensity at high resolution with the ability to capture self-consistently the evolution of the intergalactic UV background from sources (i.e. galaxies and BHs) to sinks (i.e. neutral gas in the IGM).
While modelling the assembly of a protocluster, the \Obelisk simulation resolves haloes down to the atomic cooling limit, therefore capturing the bulk of the potential sources of ionizing photons in this volume.
We have presented in some detail the improvements to the subgrid physical models we have used with respect to the previous generation of simulations such as \hagn.

In this paper, we concentrated on describing the global properties of galaxies and BHs in our simulation, and their contribution to reionization: Indeed, \Obelisk is unique in that it follows the radiation produced by both types of sources, allowing us to study directly their relative role in setting the ionization state of the Universe. We focused on an overdense region in order to probe the contribution of both types of sources in an environment where BHs are expected to make the largest contribution.
Our main results are as follows:
\begin{enumerate}
\item Stellar populations overwhelmingly dominate over the AGN as sources of reionization, and provide enough ionizing photons to complete reionization alone.  At $z \gtrsim 6$, despite the relatively higher escape fraction, AGN are responsible for less than 1\% of the total \hi photoionization rate, and represent a similarly low fraction of the total ionizing emissivity.
\item Both star formation and BH accretion are strongly enhanced in \Obelisk compared to an average environment, in good agreement with extrapolations from the protocluster population observed at $z \gtrsim 2$.
\item The delicate balance between the larger number of sources and the higher gas density leads to a reionization history close to that of an average environment ($z_{\rm reion} \sim 6$). The reionization proceeds first inside-out, with the most massive galaxies reionizing their close environment first, before the ionization fronts propagate to the voids.
\item In our protocluster environment, the global escape fractions from stars and AGN during reionization are $\fescstar \simeq 3-8\%$ and $\fescAGN \simeq 30-40\%$, respectively.
\item In high densities environments, helium double-reionization happens early, predominantly because of the large density of \heii-ionizing AGN sources compared to the field.
\end{enumerate}
The broad picture emerging from this first analysis of the \Obelisk simulation is in agreement with the traditional reionization picture, in which AGN are sub-dominant in establishing the ionizing UV background, but contribute to maintaining the Universe reionized in the post-overlap era \citep[e.g.][]{Madau1999, Haehnelt2001}. Our paper shows that this result holds even in dense regions, where the AGN contribution is expected to be enhanced,  but was here found to be still mostly irrelevant at $z \gtrsim 6$.

The results presented here are intended as an introduction to the simulation, upon which further analysis will build.
For example, we will study if the harder ionizing spectrum of AGN, which can penetrate denser gas, means that AGN play a more important role than stars in ionizing intergalactic filaments, despite the fact that they ionize only a small fraction of the volume (Fig.~\ref{fig:qhii_fractions}).
In the same spirit, we will benefit from the comparison with a twin simulation that has been run without radiative transfer to study the impact of the inhomogeneous reionization on the suppression of low-mass galaxies, in particular around quasars \citep[see e.g.][for an observational perspective]{Mazzucchelli2017a}.
On the galaxy formation side, further work is needed in conjunction with simulations of field environments (e.g. \sphinx or \newh) in order to understand how the overdensity of the \Obelisk volume affects the concurrent growth of galaxies and BHs, and if this type of environment favours AGN activity beyond the global increase in the number density of massive haloes. This type of analysis will obviously greatly benefit from the fact that \Obelisk is a sub-volume of the \hagn simulation: We will be able to connect our high-redshift results to the $z = 0$ Universe.
Finally, we will address more thoroughly the comparison of our simulated galaxies to the observed populations at high-redshift, in preparation for surveys with the upcoming JWST, as well as with ground based instrument such as MUSE on the VLT or ALMA. This will of course benefit from the inclusion of a model for dust, allowing us for instance to weigh the contributions of obscured vs. UV-bright star formation.

\begin{acknowledgements}
  We would like to thank the referee for their useful comments which have strongly improved the readability of this paper.
  We would like to thank Rebekka Bieri, Jérémy Blaizot, Sam Geen, Mélanie Habouzit, Sandrine Lescaudron and Pierre Ocvirk for helpful comments and stimulating discussions on this work, and over the inception of the \Obelisk project in general.
  
  This work made use of v2.2.1 of the Binary Population and Spectral Synthesis (\bpass) models as described in \citet{Eldridge2017} and \citet{Stanway2018}. We also used the models produced by \citet{Kulkarni2019code}.
  MT and MV acknowledge funding from the European Research Council under the European Community's Seventh Framework Programme (FP7/2007-2013 Grant Agreement no. 614199, project `BLACK'). MT is supported by Deutsche Forschungsgemeinschaft (DFG, German Research Foundation) under Germany's Excellence Strategy EXC-2181/1 - 390900948 (the Heidelberg STRUCTURES Cluster of Excellence). HP is indebted to the Danish National Research Foundation (DNRF132) and the Hong Kong government (GRF grant HKU27305119) for support. TK was supported in part by the Yonsei University Future-leading Research Initiative (RMS2-2019-22-0216) and in part by the National Research Foundation of Korea (No. 2017R1A5A1070354 and No. 2018R1C1B5036146).
  CC has received partial funding from the Institut Lagrange Paris (ILP) and is supported by the European Union’s Horizon 2020 research and innovation programme (under grant agreement No. 818085 GMGalaxies). 
  This work has made use of the Horizon Cluster hosted by Institut d'Astrophysique de Paris; we thank St{\'e}phane Rouberol for running smoothly this cluster for us. We acknowledge PRACE for awarding us access to Joliot Curie at GENCI@CEA, France, which was used to run most of the simulations presented in this work. Additionally, this work was granted access to the HPC resources of CINES under allocations A0040406955 and A0040407637 made by GENCI. This work has made extensive use of the \textsc{Yt} analysis package \citet{Turk2011} and NASA's Astrophysics Data System, as well as the \textsc{Matplotlib} \citet{Hunter2007}, \textsc{Numpy/Scipy} \citet{Jones2001} and \textsc{IPython} \citet{Perez2007} packages.
\end{acknowledgements}

\bibliographystyle{aa} 
\bibliography{obelisk.bib} 

\begin{appendix}
  \section{Effect of the low threshold for the dynamical friction}
  \label{sec:dfthreshold}

  We now investigate the effect of our very low threshold for including the dynamical friction, $\rho_{\rm DF, th}$.
  We first recall that in order to follow BH dynamics correctly, simulations are in principle required to resolve the influence radius of a the BH. Because we cannot do so in a fully cosmological context, we have to rely to a subgrid model to account for unresolved dynamical friction from the gas onto the BH. This model will tend to align the velocity of the BH to that of the gas.
  In order to account for the unresolved structure of the gas in the close vicinity of the BH, we chose to boost the frictional force by an ad hoc factor $\alpha  = (\rho/\rho_{\rm DF, th})^2$ if $\rho > \rho_{\rm th}$, the choice of the threshold $\rho_{\rm DF, th}$ controls the strength of the frictional force being arbitrary.
  
  Here, we have used a very low value for this threshold: The main effect is to increase the dynamical friction significantly, effectively sticking the BH to the gas cloud it is embedded in. We stress that this is at most comparable to the effect of artificially redirecting the BH towards the centre of the cloud. Nevertheless, we want to quantify the effect of this on the BH population in \Obelisk.
  To do so, we make use of a twin simulation that includes the exact same physical model as \Obelisk except for the radiation hydrodynamics and that will be presented in another paper ({Cadiou et al. (in prep)}). Because it is significantly cheaper, we ran two versions of this simulation, varying the value of the density threshold from $\rho_{\rm DF, th} = 0.01\,\mbox{cm}^{-3}$ to $\rho_{\rm DF, th} = 10\,\mbox{cm}^{-3}$, similar to the value used in the \newh simulation.
  \begin{figure}
    \centering
    \includegraphics[width=\linewidth]{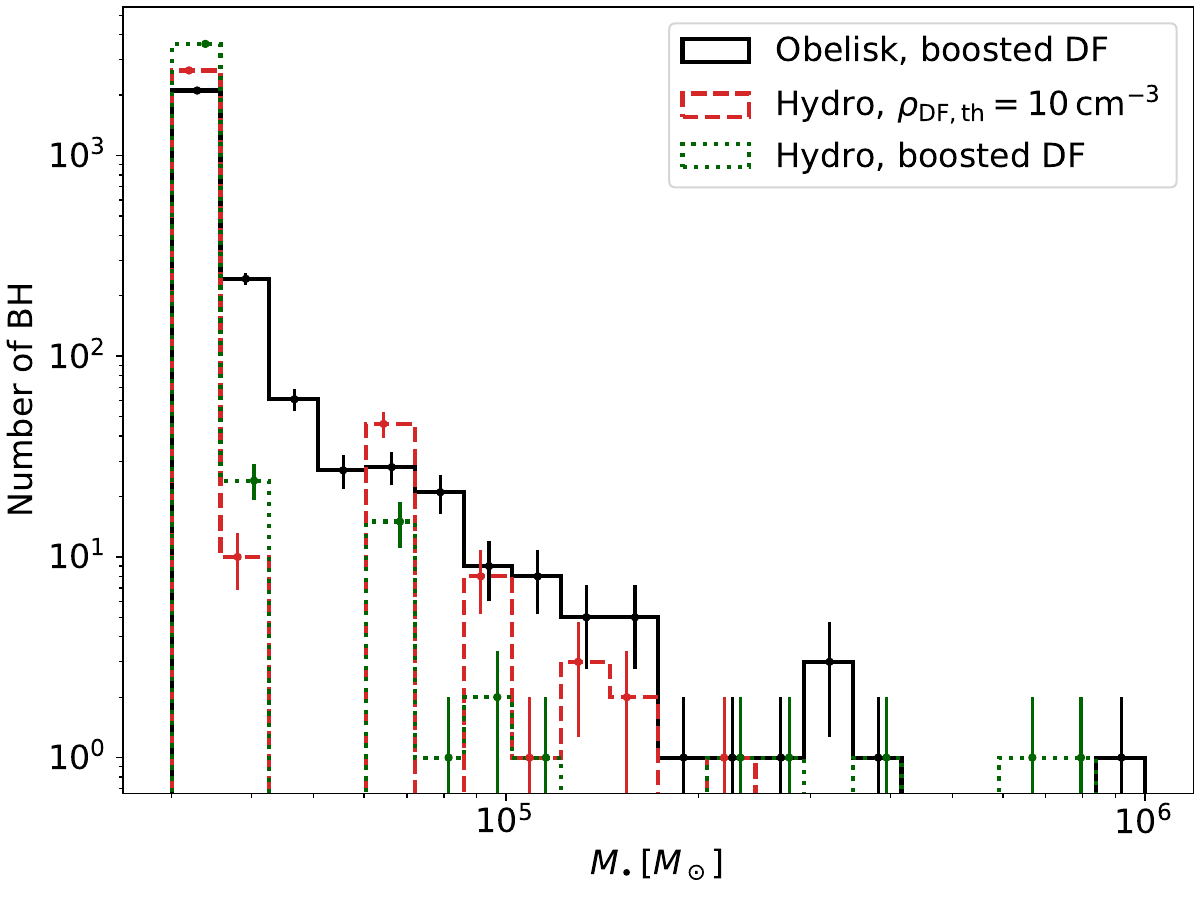}
    \includegraphics[width=\linewidth]{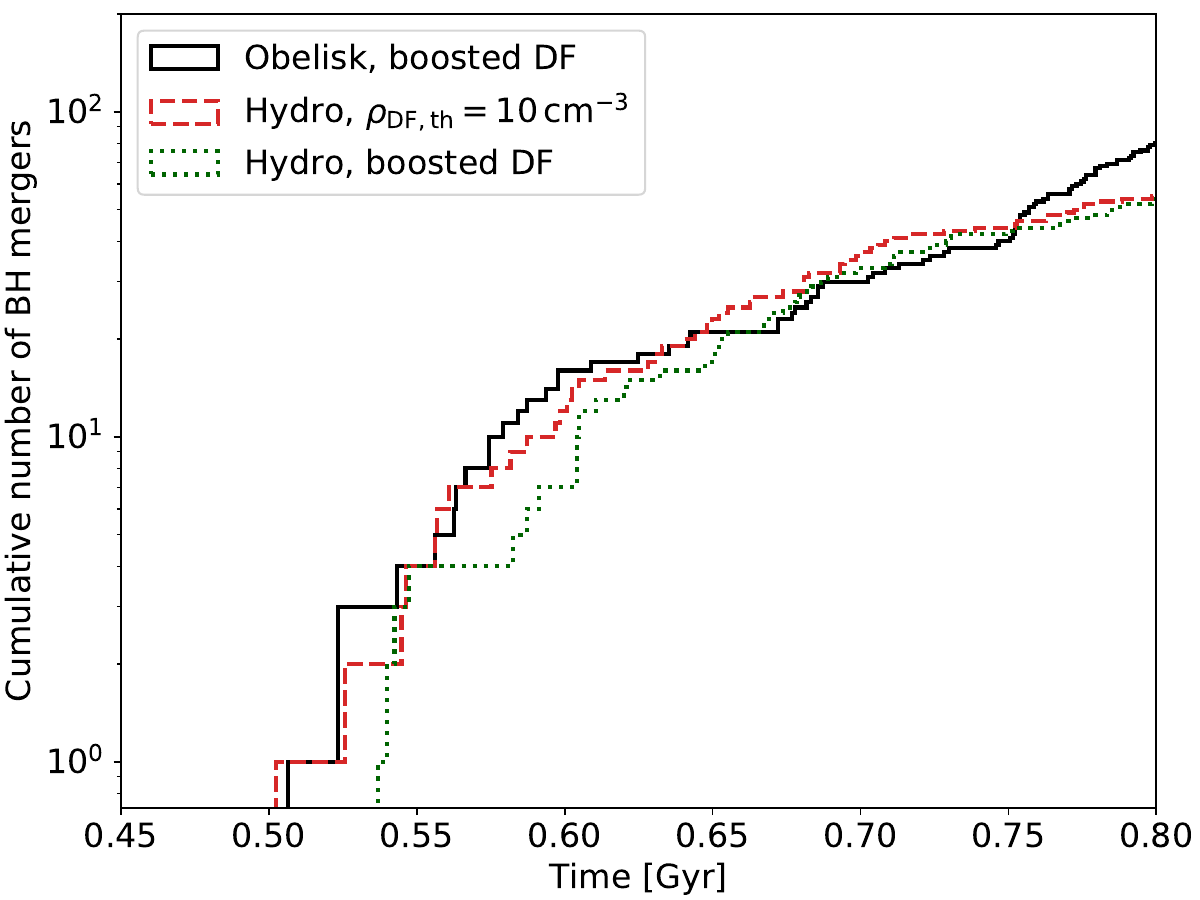}
    \caption{Distribution of BH masses (\emph{top}) and cumulative number of BH mergers (\emph{bottom}) for the fiducial RHD run (in black) and for the two twin non-RHD runs with $\rho_{\rm DF, th} = 10\,\mbox{cm}^{-3}$ (red dashed line) and with the value described in Sect.~\ref{sec:bh-dynamics} (green dotted line) at $z \sim 6.9$.}
    \label{fig:df_bhmf}
  \end{figure}
  
  We show the BH mass function at $z \sim 6.9$ for the fiducial \Obelisk simulation and for both hydro runs in the upper panel of Fig.~\ref{fig:df_bhmf}. The hydro run using the same value of $\rho_{\rm DF, th}$ as \Obelisk is show as a green dotted line and the run with the higher value corresponds to the red dashed line. The error bars indicate the $\sqrt{N}$ Poisson error in each mass bin. Comparing the two hydro runs, it seems that the run with stronger dynamical friction is marginally depleted in BHs with masses around $5\times 10^4 - 8 \times 10^4\,\Msun$, and that the most massive BH are somewhat more massive. The RHD run, with strong dynamical friction, behaves differently: There are overall more BHs at all masses, except just around the seed mass. This suggests that compared to other processes, the details of the dynamical friction from gas have a relatively minor effect, once it is strong enough to maintain the BH in the centre of the galaxy. The lower panel shows the cumulative number of BH-BH mergers measured in these three runs, with the same legend. Here, we see that in the hydro run with weaker dynamical friction experiences slightly more mergers than the other hydro run: This suggests that the main effect of our stronger dynamical friction is to slightly increase the gas density around the BH, enhancing a bit the accretion. However, we see here as well that changing the strength of the dynamical friction does not affect how BHs merge more than the inclusion of radiation hydrodynamics.
  To summarize, while our choice to strongly boost the dynamical friction seems artificial, it leads to a BH growth history that is similar to that from simulations with a weaker prescription for the BH dynamics.

  \section{AGN SED model}
  \label{sec:agn-sed-model}

 In this appendix, we present more explicitly the model for the AGN radiation already described in Sect.~\ref{sec:bh-radiation}, in order to facilitate the reproducibility of our numerical experiment.
  We highlight that our endeavour here is not to provide an AGN spectrum comparable with observations on a one-to-one basis, but rather to derive a spectral shape that will broadly capture how the ionizing luminosity of the AGN varies for different accreting BHs. Specifically, we focus on the ionizing UV bands that we consider in \Obelisk.
  In particular, we do not model the infrared emission from the dust, and only assume that a fraction $f_{\rm IR} = 30\%$ of the total AGN luminosity is absorbed by dust and re-emitted in the IR. This is close to the value suggested by the average \citet{Sazonov2004} spectrum.
  We assume that each BH for which the accretion rate exceeds $\dot{M}_{\rm BHL} \geq \chi_{\rm crit} \dot{M}_{\rm Edd} = 0.01  \dot{M}_{\rm Edd}$ is surrounded by an optically thick, geometrically thin $\alpha$-disc\footnote{Strictly speaking, this solution is in only valid if the disc luminosity stays below $0.3\,L_{\rm Edd}$ to ensure that the disc stays well described by a thin disc. We still choose to use the thin disc solution up to $L = L_{\rm Edd}$.}, as described by \citet{Shakura1973}, \citet{Novikov1973} and \citet{Page1974}.

  The emission from a column of gas located a radius $R$ in the disc can be described, under the assumption of local thermodynamical equilibrium between the gas and the radiation, as that of a blackbody of temperature $T_{\rm BB}(R)$, such that the energy flux crossing the surface of the disc $\mathcal{F}(R)$ can be equated to $\sigma_{\rm SB} T_{\rm bb}(R)^4$.
  This energy flux can be computed analytically for an $\alpha$-disc as
  \begin{equation}
    \label{eq:AGNSED_radflux}
    \mathcal{F}(R) = \frac{3 G \Mbh \dot{\Mbh}}{8\pi R^3} f\left(\frac{R}{R_{\rm g}}\right),
  \end{equation}
  where $f(r)$ is specified by the disc profile.
  For the \citet{Shakura1973} solution\footnote{The complete solution for a Kerr spacetime is given in \citet{Page1974}, but we have checked that using the \citet{Shakura1973} profile does not change significantly the resulting spectrum.}, we have
  \begin{equation}
    \label{eq:AGNSED_f_of_r}
    f(r) = f(R/R_g) = 1 - \sqrt{r_{\rm isco}/r}.
  \end{equation}
  We derive the blackbody temperature of a ring at radius $R$ as
  \begin{align}
    \label{eq:AGNSED_Tbb_of_r}
    \begin{aligned}
      T_{\rm BB}(R) &= \left(\frac{3 G \Mbh \dot{\Mbh}}{8\pi R^3} f(r)\right)^{1/4}\\
      &= \left(\frac{3 G \Mbh \dot{\Mbh}}{8\pi R_{\rm isco}^3}\right)^{1/4} \left(\frac{R}{R_{\rm isco}}\right)^{-3/4} \left(1-\sqrt{\frac{R_{\rm isco}}{R}}\right)^{1/4}.
    \end{aligned}
  \end{align}
  The ring at a radius $R$ will then emit radiation with a Planckian spectrum:
  \begin{equation}
    \label{eq:AGNSED_planck}
    B_\nu(T_{\rm BB}(R)) = \frac{2h\nu^3}{c^2} \frac{1}{e^{\frac{h\nu}{k_{\rm B} T_{\rm BB}(R)}} - 1}\,,
  \end{equation}
  with $h$ and $k_{\rm B}$ the Planck and Boltzmann constants, respectively.

  The total spectrum of the disc can then readily be computed by integrating Eq.~\ref{eq:AGNSED_planck} between the ISCO radius and the outer edge of the disc. We take this outer edge to be the self-gravitating radius $r_{\rm sg}$ of the disc, that is the radius at which the gravity of the central BH does not dominate anymore. \citet{Laor1989} give for a radiation-dominated thin disc:
  \begin{equation}
    \label{eq:AGNSED_rsg}
    r_{\rm sg} \simeq 2150 \alpha^{2/9} \left(\frac{\Mbh}{10^9\,\Msun}\right)^{-2/9} \dot{m}^{4/9}\,,
  \end{equation}
  where $\alpha \sim 0.1$ is the disc viscosity and $\dot{m} = \frac{\epsilon_r(0)}{\epsilon_r(a_\star)} \frac{L}{L_{\rm Edd}}$ is the reduced mass accretion rate of a BH with spin $a_\star$ and luminosity $L$ normalized to that of a non-spinning BH. We note that for a moderately luminous BH, the assumption that the radiation pressure dominates over the gas pressure can break down at a radius smaller than the self-gravitating radius. However, only the outermost regions of the disc, for which the blackbody temperature are the lowest, could be affected. These regions contributes very little to the ionizing UV radiation, and our results are consequently not strongly affected by this.

  The total luminosity of the disc at a frequency $\nu$ thus reads
  \begin{equation}
    \label{eq:AGNSED_Lnu}
    L_\nu = 2\pi\int_{r_{\rm isco}}^{r_{\rm sg}} \frac{2h\nu^3}{c^2} \frac{1}{e^{\frac{h\nu}{k_{\rm B} T_{\rm BB}(R)}} - 1} 2\pi R \mathrm{d}R\,.
  \end{equation}
  
  The spectral shape resulting from Eq.~\ref{eq:AGNSED_Lnu} can be approximated at low energy (in the Rayleigh-Jeans regime) by $L_\nu \propto \nu^2$, and by $L_\nu \propto \nu^{1/3}$ at intermediate energies, corresponding to the part of the disc where the temperature profile follows $T(R) \propto R^{-3/4}$. At high frequencies (corresponding to the innermost regions of the disc), however, the spectrum is exponentially cut off. Rather than trying to find a physically motivated approximate model for the high energy part, we choose to replace the exponential cut-off by a power law, $L_\nu \propto \nu^{-\alpha_{\rm UV}}$, with $\alpha_{\rm UV} = 1.5$, in broad agreement with the value $\alpha_{\rm UV} = -1.7\pm 0.61$ derived by \citet{Lusso2015} for a sample of high-redshift quasars.
  The fraction of the total luminosity in a given frequency interval $[\nu_{\rm min}; \nu_{\rm max}]$ can then be easily obtained by integrating Eq.~\ref{eq:AGNSED_Lnu} over the interval. For instance, the fraction of the luminosity in the first UV band considered in \Obelisk is obtained by
  \begin{equation}
    \label{eq:AGNSED_fUV1}
    f_{\mathrm{UV},1} = \int_{13.6\,\mbox{eV}/h}^{24.59\,\mbox{eV}/h} L_\nu \mathrm{d}\nu\,.
  \end{equation}
  
  This leads to a spectrum that depends on $\Mbh$, $\dot{\Mbh}$, and $a_\star$ that can be in principle readily used in \ramsesrt in the same manner as the stellar population models. For this latter case, it is necessary to integrate the spectrum on-the-fly as the average energy and the interaction cross-sections in each radiation bin can vary (up to a factor of a few for the cross-sections) as a function of the age and the metallicity of the stellar population \citep[see e.g.][appendix~B]{Rosdahl2013}.

  In this work, we have opted for a slightly different approach: we have tabulated the values of $f_{\mathrm{UV},i}$ for each bin $i=1,2,3$ and only interpolate between the tabulated values over the course of the simulation. In principle, this requires interpolation in a three-dimensional table (for the BH mass, accretion rate and spin). However, the model described here is very weakly dependent on the BH spin $a_\star$. Effectively, we checked that the variation in $f_{\mathrm{UV},i}$ as $a_\star$ varies is much weaker than when changing the \Mbh or $\dot{\Mbh}$. To limit the cost of our model, we drop the spin dependence and assume $a_\star \sim 0.7$ (corresponding to a radiative efficiency $\epsilon_r \simeq 0.1$) when computing the spectrum, leading to a simpler two-dimensional interpolation.
  Similarly, it appears that the average energy and cross-sections in each frequency interval depend very weakly not only on the spin, but also on \Mbh and $\dot{\Mbh}$. We therefore choose to fix their values to that of a non-spinning BH with mass $\Mbh = 10^7\,\Msun$ accreting at 10\% of its Eddington luminosity.

\end{appendix}

\end{document}